\documentclass[twocolumn]{aastex63}

\usepackage{amsthm}
\usepackage{amsmath}
\usepackage{hyperref}
\usepackage{float}
\usepackage{url}
\usepackage{natbib}
\usepackage{grffile}
\usepackage{gensymb}
\usepackage{bigints}
\usepackage{textcomp}
\usepackage{booktabs}
\graphicspath{{./} {../figures/}}

\definecolor{red}{rgb}{1.0,0,0}

\newcommand{\Rom}{\textsc{Romulus25}}
\newcommand{\HI}{\hbox{\rmfamily H\,{\scshape i}\;}}

\newcommand{\dn}{$D_{\rm n}4000\;$}
\newcommand{\halpha}{${\rm H}\alpha\;$}

\begin{abstract}
    We examine the quenching characteristics of $328$ isolated dwarf galaxies $\left(10^{8} < M_{\rm star}/M_\odot < 10^{10} \right)$ within the \Rom{} cosmological hydrodynamic simulation. 
    Using mock observation methods, we identify isolated dwarf galaxies with quenched star formation and make direct comparisons to the quenched fraction in the NASA Sloan Atlas (NSA). Similar to other cosmological simulations, we find a population of quenched, isolated dwarf galaxies below $M_{\rm star} < 10^{9} M_\odot$ not detected within the NSA.
    We find that the presence of massive black holes (MBHs) in \Rom{} is largely responsible for the quenched, isolated dwarfs, while isolated dwarfs without an MBH are consistent with quiescent fractions observed in the field. Quenching occurs between $z=0.5-1$, during which the available supply of star-forming gas is heated or evacuated by MBH feedback. Mergers or interactions seem to play little to no role in triggering the MBH feedback.  At low stellar masses, $M_{\rm star} \lesssim 10^{9.3} M_\odot$, quenching proceeds across several Gyr as the MBH slowly heats up gas in the central regions. At higher stellar masses, $M_{\rm star} \gtrsim 10^{9.3} M_\odot$, quenching occurs rapidly within $1$ Gyr, with the MBH evacuating gas from the central few kpc of the galaxy and driving it to the outskirts of the halo. Our results indicate the possibility of substantial star formation suppression via MBH feedback within dwarf galaxies in the field.  On the other hand, the apparent over-quenching of dwarf galaxies due to MBH suggests higher resolution and/or better modeling is required for MBHs in dwarfs, and quenched fractions offer the opportunity to constrain current models.
\end{abstract}

\begin{document}

\shortauthors{Sharma et al.}
\title{AGN quenching in simulated dwarf galaxies}

\author{Ray S. Sharma}
\affiliation{Department of Physics and Astronomy, Rutgers, The State University of New Jersey, 136 Frelinghuysen Road, Piscataway, NJ 08854, USA}

\author{Alyson M. Brooks}
\affiliation{Department of Physics and Astronomy, Rutgers, The State University of New Jersey, 136 Frelinghuysen Road, Piscataway, NJ 08854, USA}
\affiliation{Center for Computational Astrophysics, Flatiron Institute, 162 Fifth Avenue, New York, NY 10010, USA}

\author{Michael Tremmel}
\affiliation{Department of Astronomy, Yale University, 52 Hillhouse Ave, New Haven, CT 06511, USA}

\author{Jillian Bellovary}
\affiliation{Department of Physics, Queensborough Community College, City University of New York, 222--05 56th Ave, Bayside, NY 11364, USA}
\affiliation{Department of Astrophysics, American Museum of Natural History, Central Park West at 79th Street, New York, NY 10024, USA}
\affiliation{The Graduate Center, City University of New York, 365 5th Ave, New York, NY 10016, USA}

\author{Thomas R. Quinn}
\affiliation{Department of Astronomy, University of Washington, PO Box 351580, Seattle, WA 98195, USA}

\section{Introduction}

    Over the past two decades, large sky surveys have found a bimodality in galaxy properties that separate actively star-forming galaxies from quiescent galaxies as far back as $z=4$ \citep{Giallongo2005,Brammer2009,Willmer2006,Cassata2008,Muzzin2013,Ilbert2013}. Galaxies occupy distinct regions in color-magnitude (or similarly color-color, star formation rate-mass) space, where star-forming galaxies exhibit younger, bluer stellar populations in rotation-dominated morphologies, while quiescent galaxies exhibit older, redder populations in dispersion-dominated morphologies \citep{Strateva2001,Kauffmann2003,Bell2003,Baldry2004,Balogh2004,Baldry2006,Wyder2007,Gallazzi2008,Ilbert2010,McGee2011,Wetzel2012,LopezFernandez2018}. 
    
    Observations by \citet{Peng2010} have shown that galaxies may cease star formation through two distinct quenching pathways: through environmental quenching \citep{Kauffmann2004,Baldry2004,Bahe2015} via galaxy harrassment \citep{Feldmann2010}, tidal stripping, or ram-pressure stripping \citep{Tonnesen2007,Weinmann2009}; or through mass/internal quenching \citep{Kauffmann2003a} via, e.g, halo quenching in massive halos \citep{Rees1977,Kauffmann1993} or starvation \citep{Larson1974,Dekel1986} in both low- and high-mass halos. Internal quenching mechanisms also include the physical removal or heating of star-forming gas via feedback from stars and active galactic nuclei \citep[AGN; e.g,][]{Choi2015,Choi2018}. The standard paradigm explains that the low-mass end of the galaxy stellar mass function is suppressed by reonization \citep{Efstathiou1992,Okamoto2008,Fitts2017,Rey2019a,Rey2020,Applebaum2021,Munshi2021} and supernova (SN) feedback \citep{Larson1974,Dekel1986,Mori2002}, while the high-mass end is regulated by AGN feedback \citep{Binney1995,DiMatteo2005,Bower2006,Croton2006,Sijacki2007,Cattaneo2008}.
    
    Among low-mass galaxies in particular, it is well-known that high-density environments may quench star formation through environmental quenching processes \citep[e.g,][though, see \citet{Emerick2016}]{vandenBosch2008,Smith2012,Wetzel2013,Wetzel2014,Weisz2015,Bahe2015,Tinker2017,Simpson2018,Akins2021,Samuel2022}, and indeed quenched dwarf galaxies are thought to be exceedingly rare in the field \citep{Geha2012}. Only in recent years have internal processes been observed to quench dwarf galaxies in isolation \citep{Janz2017}. It is worth noting that among the lowest mass dwarfs $\left( M_{\rm vir} \lesssim 10^{8} M_\odot \right)$, there is observational evidence that cosmic reionization may have quenched low-mass dwarfs around $z\sim6$ \citep[e.g,]{Weisz2014,Brown2014,Bettinelli2018}, though other work finds no such signal \cite{Monelli2010,Hidalgo2011,Skillman2017}.
    
    Observational evidence is building that suggests AGN feedback may impact star formation in dwarf galaxies \citep{Silk2017}. Using MaNGA observations of quenched galaxies with stellar masses $M_{\rm star} < 5\times10^{9} M_\odot$, \citet{Penny2018} identify six quenched galaxies that exhibit offsets between star and gas kinematics alongside strong AGN-like ionization, indicative of star formation suppression driven by AGN feedback. \citet{Bradford2018} use unresolved \HI measurements to identify a population of $10^{9.2} < M_{\rm star}/M_\odot < 10^{9.5}$ galaxies with quiescent star formation, gas depletion, and strong AGN-like ionization. \citet{Manzano-King2019} present the first direct detection of high velocity, AGN-driven ionized gas outflows within a sample of $13$ optically-selected dwarf galaxies with $10^{8.78} < M_{\rm star}/M_\odot < 10^{9.95}$. Of these galaxies, six have AGN-like line ratios alongside outflow velocities exceeding the escape velocity of the host halo. \citet{Dickey2019} perform follow-up optical spectroscopy of the lowest-mass quiescent dwarfs known to exist in isolation, finding that $16$ of the $20$ observed quiescent dwarfs show signs of strong AGN-like line ratios. They find that the AGN fraction among isolated, quiescent galaxies is significantly higher than for quiescent dwarfs in group environments.

    Some simulations and analytic models of low-mass galaxies have also begun to find that AGN feedback may have a greater impact on star formation than previously thought. By assuming standard MBH scaling relations, \citet{Dashyan2018} use analytic models to show that feedback from AGN may be even more successful than SNe in driving out gas from galaxies below a critical halo mass threshold. Using a suite of cosmological simulations run with various MBH and SN feedback models, \citet{Barai2019} find that gas-rich dwarf galaxies $(10^{5} < M_{\rm star}/M_\odot < 10^{8}$ at $z=4)$ can quench at $z=5-9$ from early AGN feedback, and only the hosts of the most massive MBHs show signatures of AGN-driven outflows. \citet{Sharma2020} explores the impact of MBHs in isolated $10^{8} < M_{\rm star}/M_\odot < 10^{10}$ dwarfs in \Rom{}, finding that efficiently accreting MBHs are more likely than inefficient accretors to be found within diffuse, gas depleted, quiescent galaxies. Within the {\sc FABLE} cosmological simulations, \citet{Koudmani2021} find that over-massive (relative to the BH mass -- M$_{\rm star}$ relation) MBHs in $M_{\rm star} < 10^{9.5} M_\odot$ galaxies are able to drive hotter, faster outflows which drive material out of the host halo. They find that outflows from over-massive MBHs are strong enough to suppress star formation at high redshift, driving misalignment of the stellar and ionized gas kinematics. At $z < 2$, their AGN exhibit lower accretion rates on average, and SN feedback becomes the more dominant driver of star formation suppression. Using a suite of cosmological zoom simulations of a $M_{\rm vir}\left(z=0\right) \sim 10^{10} M_\odot$ halo, \citet{Koudmani2022} find that feedback from over-massive MBHs may regulate star formation with effects ranging from minor suppression to catastrophic quenching. Their MBHs are capable of driving ejective outflows that deplete (or partially deplete) the star-forming gas reservoir. Quiescence is maintained through the AGN heating the circumgalactic medium and suppressing future cold gas accretion into the galaxy.
    
    On the other hand, some simulations find that MBH feedback is sub-dominant to SN feedback in impacting star formation. \citet{Habouzit2017} find that SN-driven winds within low-mass {\sc SuperChunky} halos $\left(M_{\rm halo} \lesssim 10^{11} M_\odot\right)$ can remove dense, star-forming gas, and inhibit gas accretion onto the central MBH. \citet{Koudmani2019} test various AGN and SN feedback models in high-resolution simulations of a $M_{\rm star} \sim 2 \times 10^{9} M_\odot$ dwarf galaxy. They find that AGN can drive hotter, higher velocity outflows in-line with results from \citet{Penny2018}, but alone are unable to drive global changes in star formation. In their simulations, it is the combination of AGN feedback and SN feedback that can impact star formation.
    
    Cosmological simulations have only in recent years achieved the resolutions and complexity to be able to accurately study quenching in galaxies at Milky Way masses and below \citep[e.g,][]{Kimm2015,Sanchez2021,Joshi2021,Davies2022}. In the past, simulations have struggled to generate consistent fractions of quenched satellite galaxies \citep{Font2008,Kimm2009,Weinmann2006,Hirschmann2014,Somerville2015}. Now, simulations can better reproduce the observed bimodal distributions in star formation rate \citep{Feldmann2017}, morphology \citep{Snyder2015a}, and color \citep{Nelson2018, Akins2021}. However, \citet{Dickey2021} find that large-scale cosmological simulations tend to over-predict the numbers of isolated, quiescent galaxies between $10^{8} < M_{\rm star}/M_\odot < 10^{10}$. Even when accounting for observational and selection biases as well as differences in resolution, cosmological simulations tend to quench dwarf galaxies more often than observed in the optical.
        
    In this work, we analyze the characteristics of isolated, quiescent dwarf galaxies in the \Rom{} cosmological hydrodynamical simulation. We select our sample using the mock observation process introduced in \citet{Dickey2021}, allowing us to make direct comparisons with optical observations and other mock-observed simulations. In Section \ref{Methods} we outline the relevant physics in the simulation, as well as the mock observation procedure used in this work. In Section \ref{Results}, we explore the differences in star formation histories, stellar morphology, gas content, and gas evolution between quiescent and star-forming dwarfs between $10^{8} < M_{\rm star}/M_\odot < 10^{10}$. We further investigate the quenching mechanism in our isolated dwarfs, assessing the impact of both environment and feedback effects on star formation suppression. \Rom{} is well-equipped for this analysis as one of few large-scale cosmological simulations capable of resolving the growth and dynamics of MBHs in low-mass galaxies, allowing us to accurately assess their importance to low-mass galaxy quenching.
    
\section{Methods} \label{Methods}

    \subsection{Simulation}

        Here we summarize the relevant aspects of the \Rom{} cosmological hydrodynamical simulation. The full description of the physical prescriptions can be found in \citet{Wadsley2017,Tremmel2017}.
        
        The \textsc{Romulus} suite of cosmological hydrodynamic simulations were run with the Tree+SPH code \textsc{ChaNGa} \citep{Menon2015, Wadsley2017}. In the \Rom{} ($25$ Mpc)$^{3}$ uniform-resolution volume, dark matter particles have masses of $3.39 \times 10^{5} M_\odot$ while gas particles have masses of $2.12 \times 10^{5} M_\odot$. Particles have a Plummer equivalent force softening of $250$ pc. \Rom{} contains $3.375\times$ more dark matter particles than gas in order to better resolve the dynamics of MBHs \citep{Tremmel2015}. \Rom{} uses a Planck 2014 $\Lambda$CDM cosmology, with $\Omega_m = 0.3086$, $\Omega_\Lambda = 0.6914$, $h = 0.6777$, and $\sigma_8 = 0.8288$, \citep{PlanckCollaboration2014}.
    
        Star formation is governed by the star formation efficiency, SN feedback efficiency, and the gas density threshold for star formation. Star formation follows a \citet{Kroupa2001} initial mass function. These free parameters were tuned in \citet{Tremmel2017} with a suite of $80$ zoom simulations of halos with halo masses $M_{\rm vir} = 10^{10.5} - 10^{12} M_\odot$. Star formation parameters were chosen by their ability to reproduce local scaling relationships between stellar mass, halo mass, \HI gas mass, and angular momentum \citep{Moster2013, Obreschkow2014, Cannon2011, Haynes2011}, resulting in the following star formation parameters:
        \begin{itemize}
            \item Star formation efficiency $c_\ast = 0.15$,
            \item SN feedback efficiency $\epsilon_{\rm SN} = 0.75$,
            \item Gas density threshold $n_\ast = 0.2$ m$_p$ cm$^{-3}$.
        \end{itemize}
        
        \Rom{} also contains models for metal diffusion \citep{Shen2010}, thermal diffusion \citep{Governato2015}, and low-temperature radiative cooling \citep{Guedes2011}.
        
        MBHs are seeded similar to a direct-collapse model \citep{Haiman2013,Greene2020}, using properties of local, pre-collapse gas. Gas particles form MBH seeds if they follow these criteria:
        \begin{itemize}
            \item Low metallicity, $Z < 3\times10^{-4}$,
            \item Gas density threshold, $n_\ast > 3$ m$_p$ cm$^{-3}$,
            \item Temperature, $T \sim 10^{4}$ K,
        \end{itemize}
        which were chosen to restrict BH seeding to form gas that is both collapsing rapidly and cooling slowly. These criteria approximate the seeding environments from theoretical estimates \citep{Begelman2006,Volonteri2012}.
        Once the gas particle meets these criteria, it seeds an MBH with mass $M_{\rm BH} = 10^{6} M_\odot$, gathering mass from nearby gas particles if necessary. These criteria restrict nearly all MBH seeding to above redshift $z \gtrsim 5$.
        
        Dynamical friction prescriptions are employed in \Rom{} in order to realistically model MBH sinking timescales \citep{Tremmel2015}. Dynamical friction sub-grid models allow MBHs in large halos to stay centered \citep{Kazantzidis2005,Pfister2017} without forced MBH re-centering. This prescription also allows for ``wandering'' MBHs, which move more freely throughout shallow potentials, as has been discovered in recent work \citep{Bellovary2019,Reines2020,Ricarte2021,Ricarte2021a}. The oversampling of dark matter, the high MBH seed mass, and the dynamical friction prescription together mitigate unrealistic numerical heating of MBHs, ensuring accurate MBH dynamics.

        Feedback from MBHs follows a modified blastwave SN feedback model \citet{Stinson2006}, where thermal energy is isotropically injected into the $32$ nearest gas particles within some time, $dt$, following:
        \begin{align}
            E_{\rm BH} = \epsilon_r \epsilon_f \dot{M} c^2 dt,
        \end{align}
        where $E_{\rm BH}$ is the injected thermal energy, $\epsilon_r = 0.1$ the radiative efficiency, $\epsilon_f = 0.02$ the energy injection efficiency, and $\dot{M}$ the MBH accretion rate. MBH accretion in \Rom{} follows a modified Bondi-Hoyle prescription that accounts for angular momentum on resolved spatial scales, while enforcing Eddington-limited accretion. Accretion is handled on the smallest simulation time resolution element, typically $10^{4} - 10^{5}$ yr.
        
        The free parameters for MBH accretion and feedback were tuned through an additional $48$ zoom simulations of the same halos used for star formation parameter tuning. Instead, the star formation parameters were fixed while the MBH parameters were allowed to vary in order to match the observed $z=0$ BH mass - stellar mass relation \citep{Schramm2013}.
        
        In this work, we use the Amiga Halo Finder \citep{Knollmann2009} to extract gravitationally bound dark matter halos, sub-halos, and their corresponding baryonic content. AHF defines the virial radius, $R_{\rm vir}$, and virial mass, $M_{\rm vir}$, using a spherical top-hat collapse technique. Each halo center is estimated using a shrinking spheres approach \citep{Power2003}. When discussing halo masses, we define
        \begin{align}
            M_{\rm 200} = 4 \pi \rho_{\rm c} \times 200 R_{\rm 200}^3,
        \end{align}
        where $\rho_{\rm c}$ is the cosmic critical density and $R_{\rm 200}$ is the halo radius.

    \subsection{Mock Observations}
        
        In order to make direct comparisons of simulation results with observations, we use the mock observation package {\it orchard}\,\footnote{https://github.com/IQcollaboratory/orchard} \citep{Dickey2021}. The {\it orchard} package was designed to allow direct comparisons of galaxy quenched fractions between uniform cosmological simulations and observations. By adding realistic noise and observational limits to synthetic spectra, {\it orchard} accounts for the effects of observational incompleteness, finite signal-to-noise, and imprecise distance information for galaxies in a wide-field survey. In the following sections, we outline the mock-observation procedure.
        
        \subsubsection{Synthetic Spectra}
        
            Following \citet{Dickey2021}, we first generate synthetic spectra for each target galaxy using stellar population synthesis models from {\sc FSPS} \citep{Conroy2009,Conroy2010} with the Python interface {\it python-fsps} \citep{DanForeman-Mackey2014}. When generating spectra, we use a \citet{Chabrier2003} initial mass function to match the assumed IMF from \citet{Dickey2021}. For each galaxy at $z=0.05$, we bin the total stellar mass across a uniform grid of formation age and stellar metallicity. Formation age bins increase linearly with lookback time, $\tau$, between $\tau=\left[0, 13.1\right]$ Gyr. Metallicity bins run between $\log_{\rm 10}(Z/Z_\odot) = \left[-2.2, 0.5\right]$ with bin size $\Delta Z = 0.3$ at low metallicity, increasing to $\Delta Z = 0.1$ near $Z=0$. To avoid resolution effects on the most recent star formation, we calculate star formation rates (SFRs) younger than $15$ Myr using the expected instantaneous SFR from the gas particles that meet the temperature and density criteria necessary to form stars.
            
            Nebular emission lines are calculated by {\sc FSPS} within each metallicity bin using a {\sc CLOUDY} lookup table \citep{Byler2017} with gas metallicity set to the bin metallicity. We also include the effects of dust on our synthetic spectra. We use the \citet{Villaume2015} dust emission model for asymptotic giant branch stars. We use the two-component dust model from \citet{Charlot2000}, with power-law dust attenuation index $\Gamma=0.7$ and normalization $\tau(5500 {\rm \AA}) = 0.33$. Stars younger than $30$ Myr have additional attenuation, with index $\Gamma=0.7$ and normalization $\tau(5500 {\rm \AA}) = 1.0$. To produce a galaxy spectrum, we sum the spectra of all stellar populations in each bin of the $t, Z$ grid, weighting by the stellar mass formed.

        \subsubsection{Mock Survey}
        
            \begin{figure}
                \plotone{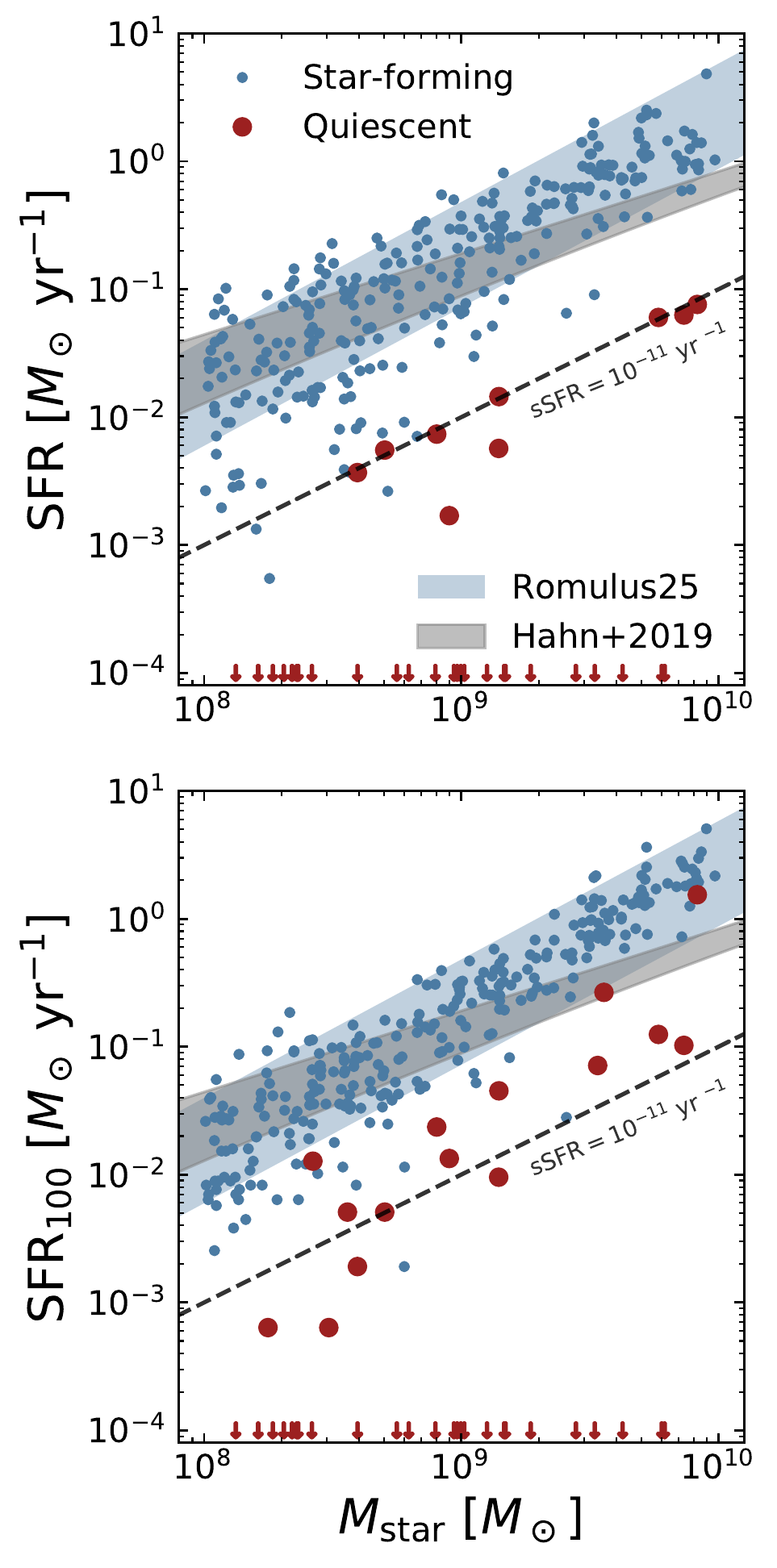}
                \caption{The instantaneous (top) and $100$ Myr averaged (bottom) intrinsic SFR versus stellar mass for isolated dwarfs at $z=0.05$. Quiescent dwarfs (red circles) selected through \halpha EW and \dn exhibit lower SFRs at fixed stellar mass than star-forming dwarfs (blue dots). The SFR$-M_{\rm star}$ relation derived from star-forming dwarfs (blue shaded interquartile range) approximately agrees with the main sequence found from SDSS \citep{Hahn2019}, though simulation SFRs are slightly over-predicted at the highest masses shown. $64\%$ of isolated, quiescent dwarfs in \Rom{} have $100$ Myr averaged SFR below $10^{-4} M_\odot$ yr$^{-1}$.}
                \label{sfr-mstar}
            \end{figure}
        
            We next generate a mock survey emulating SDSS observing characteristics using basic halo properties along with the synthetic spectra. The detailed steps are outlined in \citet{Dickey2021}, which we summarize as follows:
            
            \begin{enumerate}
                \item Select a random location $10$ Mpc outside the simulation volume to place a mock observer.
                \item Calculate apparent SDSS $g$- and $r$-band magnitudes, radial velocity, and projected distance for all galaxies as the observer would see them on the sky.
                \item Convolve each noise-free, synthetic spectrum with the SDSS instrumental line-spread profile, and resample to match the SDSS wavelength resolution and coverage. We model the line-spread profile as a Gaussian with $\sigma = 70$ km s$^{-1}$.
                \item Add SDSS-like noise to each spectrum in the mock observer's frame, using the average S/N at each wavelength found by \citet{Dickey2021}. Noise is added based on galaxy color, apparent magnitude, and wavelength by drawing from a normal distribution $\sigma(\lambda, g-r, m_r)$.
                \item Measure \dn \citep{Balogh1999} and \halpha equivalent width (EW) \citep{Yan2006} from the noise-added, instrumentally broadened spectra.
            \end{enumerate}
            
            From these noise-added measurements of \dn and \halpha EW, we are able to identify quiescent galaxies within \Rom{} as an observer might, along a given line of sight. The \halpha EW traces star formation within the past $10$ Myr, while \dn probes long-term star formation over the past $1$ Gyr by quantifying the strength of the $4000$\AA\,break in each noise-added spectrum \citep{Balogh1999,Brinchmann2004}. We consider a galaxy along a given sight-line to be quiescent if ${\rm H\alpha\, EW} < 2$\AA\,and $D_{\rm n}4000 > 0.6 + 0.1\log_{\rm 10}M_{\rm star}$ \citep{Geha2012}.
            
            Figure \ref{sfr-mstar} shows the relationship between intrinsic SFR and $M_{\rm star}$ among the isolated dwarf galaxies in \Rom{} at $z=0.05$ (see Section \ref{Results} for further explanation). In the top panel, we calculate instantaneous SFR using the expected SFR from gas particles meeting the temperature and density thresholds for star formation. In the bottom panel, we average the SFR over the past $100$ Myr. We distinguish between galaxies identified by {\it orchard} as quiescent (red circles) versus star-forming (blue circles), where quiescence is determined indirectly from the noise-added \halpha EW and \dn for a single sight-line. As in \citet{Dickey2021}, we only show galaxies brighter than $r$-band apparent magnitude $m_r < 17.7$ and $r$-band surface brightness $\mu_r < 23$ mag arcsec$^{-2}$.
            
            Following \citet{Ricarte2019}, we calculate the star formation main sequence in \Rom{} by fitting the median SFR of star-forming galaxies in $0.1$ dex stellar mass bins between $10^{8} < M_{\rm star}/M_\odot < 10^{10}$, estimating the uncertainty with interquartile ranges.
            We calculate both the instantaneous SFR and $100$ Myr averaged SFR from the simulation to better compare with the SDSS-derived SFRs found in \citet{Hahn2019}. They combine the SDSS DR7 massive, central galaxy sample from \citet{Tinker2011} with the NASA Sloan Atlas low-mass, isolated galaxy sample from \citet{Geha2012}. 
            Their sample spans stellar masses $10^{8} M_\odot \lesssim M_{\rm star} \lesssim 10^{12} M_\odot$ at $z\sim0$ (see \citet{Hahn2019} for more detail). The observations exhibit a slope of SFR$\,\propto M_{\rm star}^{0.69}$. Isolated dwarf galaxies in \Rom{} exhibit a similar main sequence to the observations down to $M_{\rm star} = 10^{8} M_\odot$, albeit with a steeper slope, SFR$\,\propto M_{\rm star}^{1.08}$ for our sample.
            
            Because our quiescent definition probes the luminosity-weighted stellar age rather than the SFR directly, there are cases of quiescent galaxies with higher average SFRs than star-forming galaxies at a given stellar mass. Regardless, we find that galaxies indirectly identified as quiescent typically fall below the main sequence, with $63\%$ of quiescent dwarfs exhibiting SFR $<10^{-4} M_\odot$ yr$^{-1}$ no matter how SFR is derived. Indirect identification of quiescent galaxies also tends to identify dwarfs below the commonly used specific star formation rate indicator of quiescence ${\rm sSFR} = 10^{-11}$ yr$^{-1}$ (black dashed), \citep[e.g,][]{Wetzel2014}. This tendency is particularly true when using instantaneous sSFRs, where $93\%$ of quiescent dwarfs fall below the instantaneous sSFR threshold, versus the $73\%$ of quiescent dwarfs that fall below the $100$ Myr threshold. Overall, the \Rom{} quiescent fractions shown in Section \ref{quiescent-fractions} change little when defining quiescence through either cuts on sSFR, cuts on distance from the main sequence, or through {\it orchard}.
    
\section{Results} \label{Results}
    
    In order to analyze quenching properties of simulated galaxies and compare to observations, we must first make a correction to the simulated stellar masses. By comparing the stellar-to-halo mass relation from high resolution hydrodynamical simulations to abundance matching estimates, \citet{Munshi2013} find that stellar masses derived from aperture photometry tend to systematically underestimate the ``true'' stellar mass from simulations by around $50\%$. Similarly, \citet{Dickey2021} find that stellar masses derived from $g-r$ colors \citep{Mao2021} and \citet{Bell2003} mass-to-light ratios tend to underestimate the ``true'' stellar mass by $\sim0.3$ dex. Previous studies of \Rom{} \citep[e.g,][]{Ricarte2019,Sharma2020} have typically used stellar mass corrections from \citet{Munshi2013}. In order to stay consistent with results from \citet{Dickey2021}, we calculate stellar masses throughout this work using our measured $g$- and $r$-band magnitudes according to \citet{Mao2021}:
    \begin{align}
        \log \left( M_{\rm star}/M_\odot\right) = 1.254 + 1.098 \left(g - r\right) - 0.4 M_{r},
    \end{align}
    where $M_{r}$ is the absolute $r$-band magnitude, assuming an absolute solar $r$-band AB magnitude of $4.64$ \citep{Conroy2009,Conroy2010}. We confirm that the stellar masses derived using this approach are consistent (within $0.2$ dex) with the adjusted stellar masses from previous works using \textsc{Romulus25}. Previous works incorporated a flat $60\%$ correction to the raw simulated stellar masses, as motivated by results from \citet{Munshi2013}. Regardless, we find that the choice of stellar mass correction has no impact on measured quiescent fractions.
    
    In this work we are primarily interested in dwarf galaxies between stellar masses $10^{8} < M_{\rm star}/M_\odot < 10^{10}$ at $z=0.05$. This mass range is just below the regime where feedback is often thought to change from SN-dominated to AGN-dominated \citep[e.g,][]{Habouzit2017}, as well as the regime where the MBH occupation fraction reaches unity in \Rom{} \citep{Sharma2022}. 
    
    In addition to selecting galaxies with low masses, we also set criteria on the galaxy local environment. We select isolated galaxies following \citet{Geha2012} using cuts in radial velocity and projected separation along the same single sight-line. For each target galaxy, we first identify potential massive neighbors with $M_{\rm star} > 2.5 \times 10^{10} M_\odot$ within an upper bound of $7$ Mpc projected distance and relative radial velocity $v_{\rm r} <1000$ km s$^{-1}$. We consider galaxies which have no massive neighbors within $1.5$ projected Mpc $(d_{\rm host} < 1.5$ Mpc$)$ to be isolated. Observations have found that galaxies with $M_{\rm star} \lesssim 10^{9} M_\odot$ are nearly all star-forming at this distance from a massive galaxy \citep{Geha2012,Rasmussen2012,Penny2016}. By following the isolation criteria, we mitigate the possibility of quenching from environmental effects (however, see Section \ref{feedback} for an analysis of the possibility of splashback galaxies).
    
    There are times in our analysis where we distinguish between dwarf galaxies with and without central MBHs. We define central MBHs as those within $2$ kpc of the halo center. In cases where there are multiple MBHs \citep[see e.g.,][]{Tremmel2018}, we define the central MBH to be the most massive within the central $2$ kpc. Only a handful of galaxies in our sample host their most massive MBHs at large distances ($> 2$ kpc) from the center. 
    
    Overall, these restrictions yield a sample of $328$ well-resolved, isolated dwarf galaxies. Table \ref{sample-size} provides a breakdown of the sample along the single sight-line used throughout much of this work.
    
    \begin{table}[]
        \centering
        \begin{tabular}{@{}lccc@{}}
            \toprule
             & \multicolumn{1}{l}{Star-forming} & \multicolumn{1}{l}{Quiescent} & \multicolumn{1}{l}{Total} \\ \midrule
            Central MBH    & 117 & 34 & 151 \\ \midrule
            Off-center MBH & 17  & 1  & 18  \\ \midrule
            No MBH         & 152 & 7  & 159 \\ \midrule
            Total          & 286 & 42 & 328 \\ \bottomrule
        \end{tabular}
        \caption{Summary of the isolated dwarf galaxy sample along the sight-line used throughout this work. We distinguish between star-forming versus quiescent galaxies, as well as galaxies with central MBHs ($<2$ kpc), off-center MBHs ($>2$ kpc), and galaxies entirely without MBHs.}
        \label{sample-size}
    \end{table}

    \subsection{Dwarfs with MBHs are more quiescent} \label{quiescent-fractions}
    
        \begin{figure*}
            \plotone{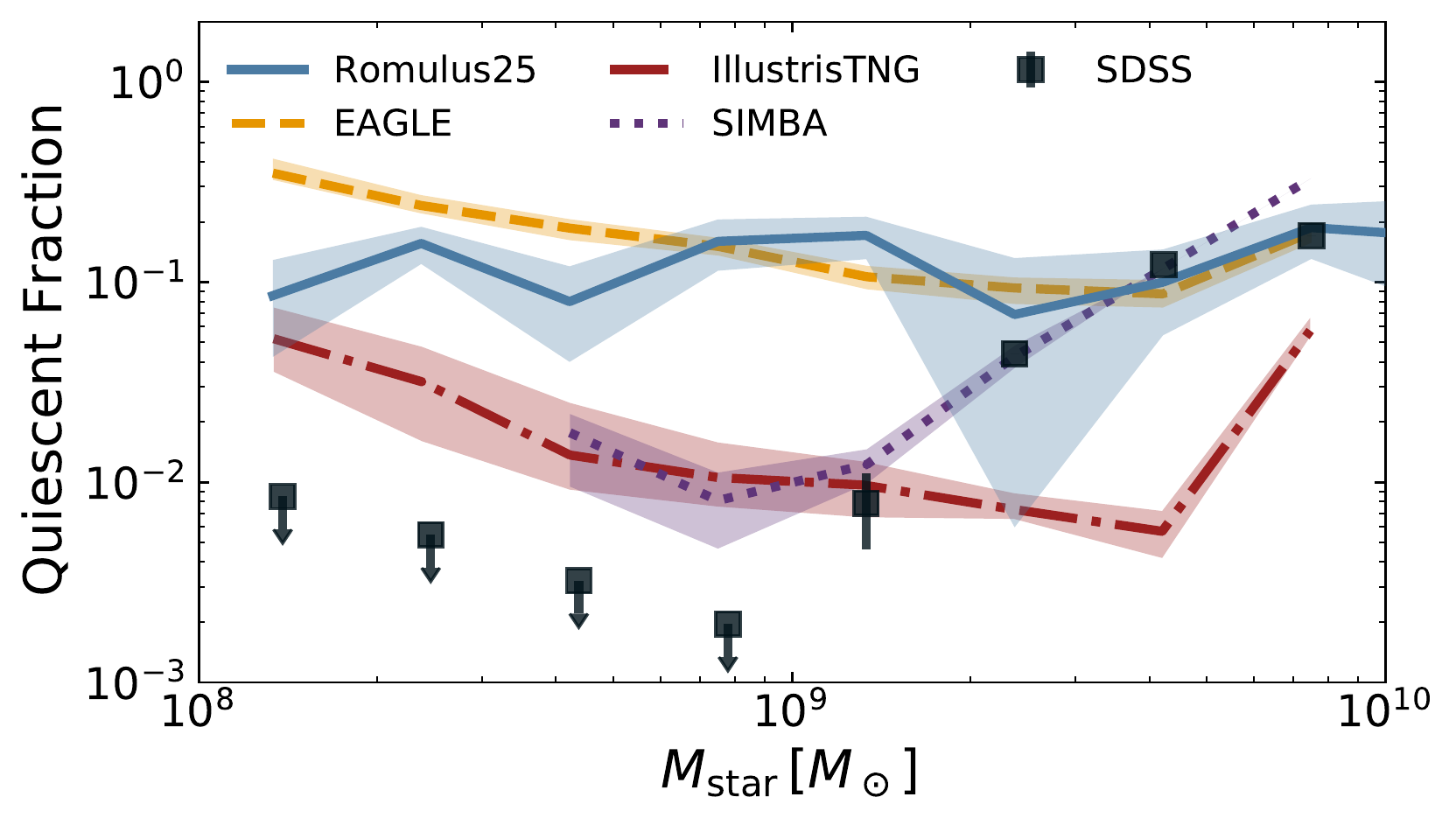}
            \caption{The quiescent fraction for isolated dwarf galaxies in \Rom{} (blue solid), {\sc EAGLE} (yellow dashed), {\sc IllustrisTNG} (red dot-dashed), {\sc SIMBA} (purple dotted), and SDSS \citep[][black squares]{Geha2012}. Each curve represents the median across $25$ randomly selected sight-lines for galaxies above a limiting apparent magnitude of $m_r=17.7$ and an $r$-band surface brightness of $\mu_r = 23$ mag arcsec$^{-2}$, with the shaded regions representing the $95\%$ binomial uncertainties averaged across all sight-lines (see text for details). All simulations except {\sc IllustrisTNG} tend to reproduce the quiescent fraction for higher mass dwarfs $M_{\rm star} > 2\times10^{9} M_\odot$, but every simulation predicts $5-15\times$ more quiescent low-mass dwarfs than found in observations.}
            \label{qfrac-comparison}
        \end{figure*}
        
        \begin{figure}
            \plotone{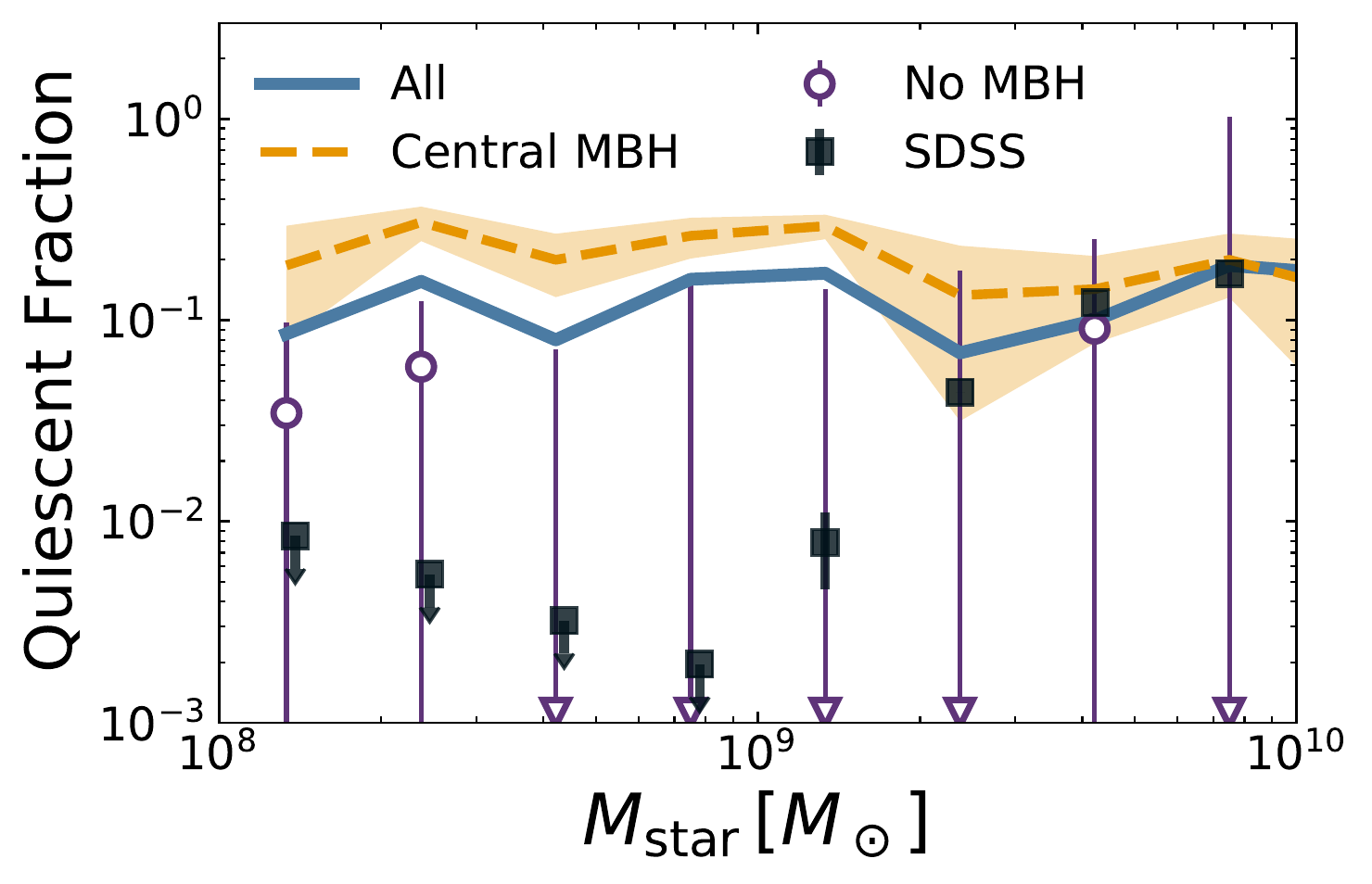}
            \caption{Median quiescent fraction across all $25$ randomly selected sight-lines (blue solid), highlighting the contribution from dwarfs with central MBH (yellow dashed) versus dwarf galaxies entirely without MBHs (purple open circle). The median quiescent fraction for dwarfs without MBHs frequently fall below the lower limits of the figure, in which case they are marked by an unfilled arrow with upper limits on the uncertainties. Dwarfs with central MBHs are $2-10\times$ more likely to be quiescent relative to dwarfs without MBHs. Dwarfs with central MBHs elevate the overall quiescent fraction for isolated dwarfs, while dwarfs without MBHs exhibit quiescent fractions consistent with those found in SDSS at M$_{\rm star} < 10^9$ M$_{\odot}$ \citep[][black squares]{Geha2012}.}
            \label{qfrac-bhs}
        \end{figure}

        We begin by examining the fraction of quiescent, isolated dwarf galaxies in \Rom{} following the methods from \citet{Dickey2021}. We repeat our mock-observation process for $25$ randomly selected sight-lines outside the simulation box, remeasuring the noise-added \dn and \halpha EW values for each galaxy and ultimately identifying detectable quiescent galaxies for each sight-line. In calculating the quiescent fraction, we weight each galaxy by the inverse of the survey volume over which the galaxy could be observed given the SDSS spectroscopic magnitude limit, $1/V_{\rm max}$. We can hence write the quiescent fraction:
        \begin{align}
            f_{\rm q} = \frac{
                \sum\limits_{i=1}^{N_{\rm q}}1/V_{\rm max, i}
            }{
                \sum\limits_{i=1}^{N_{\rm all}} 1/V_{\rm max, i}
            },
        \end{align}
        where $N_{\rm q}$ and $N_{\rm all} = N_{\rm q} + N_{\rm sf}$ are the numbers of quiescent and quiescent + star-forming galaxies in each mass bin, respectively. As in \citet{Dickey2021}, we calculate the quiescent fraction for galaxies brighter than a limiting apparent magnitude of $m_r=17.7$, and brighter than an $r$-band surface brightness of $\mu_r = 23$ mag arcsec$^{-2}$. Galaxies fainter than either limit along a given sight-line are removed from the quiescent fraction calculation.
        
        Figure \ref{qfrac-comparison} illustrates the $z=0.05$ quiescent fraction in \Rom{} as a function of stellar mass. We show the median across all $25$ sight-lines, with uncertainties reflecting the $95\%$ binomial uncertainties averaged across all $N$ sight-lines:
        \begin{align*}
            \sigma_{\rm tot}^2 = \sum_i\frac{\sigma_i^2}{N}.
        \end{align*}
        For comparison, we include the local quiescent fractions from the {\sc EAGLE} \citep{Schaye2015}, {\sc IllustrisTNG} \cite{Marinacci2018}, and {\sc SIMBA} \citep{Dave2019} simulations, as calculated by \citet{Dickey2021} using {\it orchard} with the same mock-observation procedure, quiescent definition, and stellar mass definition. We also include the observed quiescent fraction from \citet{Geha2012,Dickey2021} for mass-selected dwarf galaxies within $z < 0.055$ from the NASA/Sloan Atlas \citep{Blanton2011}.
        
        Quiescent fractions in \Rom{} agree with observations down to $M_{\rm star} > 2\times10^{9} M_\odot$, but over-predict the number of quiescent galaxies by $10-100\times$ when moving to lower masses. As discussed in \citet{Dickey2021}, all simulations (including \Rom{}) find non-zero quiescent fractions for isolated dwarfs below $M_{\rm star} < 10^{9} M_\odot$, in direct tension with SDSS observations. At $M_{\rm star} = 10^{9} M_\odot$, {\sc IllustrisTNG} and {\sc SIMBA} are closest with quiescent fractions $\sim 1\%$, while \Rom{} and {\sc EAGLE} find quiescent fractions $\sim16\%$. The weak dependence of quiescent fraction on stellar mass in \Rom{} is also most similar to {\sc EAGLE}.
        
        We now illustrate the effects of central MBHs on the quiescence of dwarf galaxies. As before, we run {\it orchard} across $25$ randomly chosen sight-lines, but now separate isolated dwarfs hosting a central MBH from those entirely without an MBH. We define central MBHs as the most massive within $2$ kpc of the halo center. Figure \ref{qfrac-bhs} shows the quiescent fraction for isolated dwarfs with central MBHs, and for those without MBHs. For comparison, we also include the observational constraints from \citep{Dickey2021}.
        
        Isolated dwarfs with central MBHs exhibit higher quiescent fractions than dwarfs without MBHs. Among dwarfs without any MBHs, the quiescent fraction is consistent with what is observed, varying between $0-8\%$. Among dwarfs with central MBHs, the quiescent fraction is substantially greater, varying between $15-30\%$. The overall quiescent fraction is largely driven by the quiescent fraction among dwarfs with MBHs. As shown in \citet{Sharma2022}, the MBH occupation in \Rom{} begins to drop below $M_{\rm star} \lesssim 10^{10} M_\odot$. Despite being more rare in dwarf galaxies, the presence of MBHs appear to correlate with quiescence down to at least $M_{\rm star} \sim 10^{8} M_\odot$.
        
        It is worth noting that the MBH seeding mechanism and occupation fractions differ between each cosmological simulation we compare with. \textsc{IllustrisTNG} forms an MBH within a halo once it reaches a mass threshold $M_{\rm halo} > 5\times10^{10} M_\odot$ \citep{Weinberger2017}. \textsc{EAGLE} similarly seeds MBHs in halos that reach a halo mass $M_{\rm halo} > 1.5\times10^{10} M_\odot$. \textsc{SIMBA} instead seeds MBHs in galaxies with $M_{\rm star} \gtrsim 10^{9.5} M_\odot$ \citep{Dave2019}. These cosmological simulations rarely (if ever) produce MBHs in halos at the lowest masses explored in this work. Thus, while our high quenched fractions at $M_{\rm star} < 10^{9} M_\odot$ appear to be connected to the presence of MBHs, MBHs are unlikely to be related to the high quenched fractions in dwarfs for other simulations.
        
        We may also ask how detectable the MBHs in our dwarf galaxies are. Results from \citet{Sharma2022} indicate that \Rom{} dwarfs often host ``hidden'' MBHS that are undetectable to current instrumentation. Such MBHs exhibit one or more of the following traits: low x-ray luminosities in the $2-10$ keV band $\left(L_{\rm X}^{\rm AGN} < 10^{39} \,\mathrm{erg\;s}^{-1}\right)$, low luminosities relative to x-ray contaminants $\left(L_{\rm X}^{\rm AGN} < 2\times L_{\rm X}^{\rm cont}\right)$, and/or large distances from the halo center $\left(d_{\rm BH} > 2\,{\rm kpc}\right)$.
        
        Applying the same methodology from \citet{Sharma2022} to the dwarfs in our sample, we find that quiescent dwarfs are more likely to host hidden MBHs than star-forming dwarfs, but the majority of MBHs are not hidden. Among quiescent dwarfs with MBHs in our sample, $31\%$ are hidden according to the criteria above, compared to the $21\%$ of star-forming dwarfs. In other words, the majority of MBHs found in our dwarf sample are detectable to current x-ray instrumentation.

    \subsection{Past growth in quiescent dwarfs}
    
        \begin{figure*}
            \plottwo{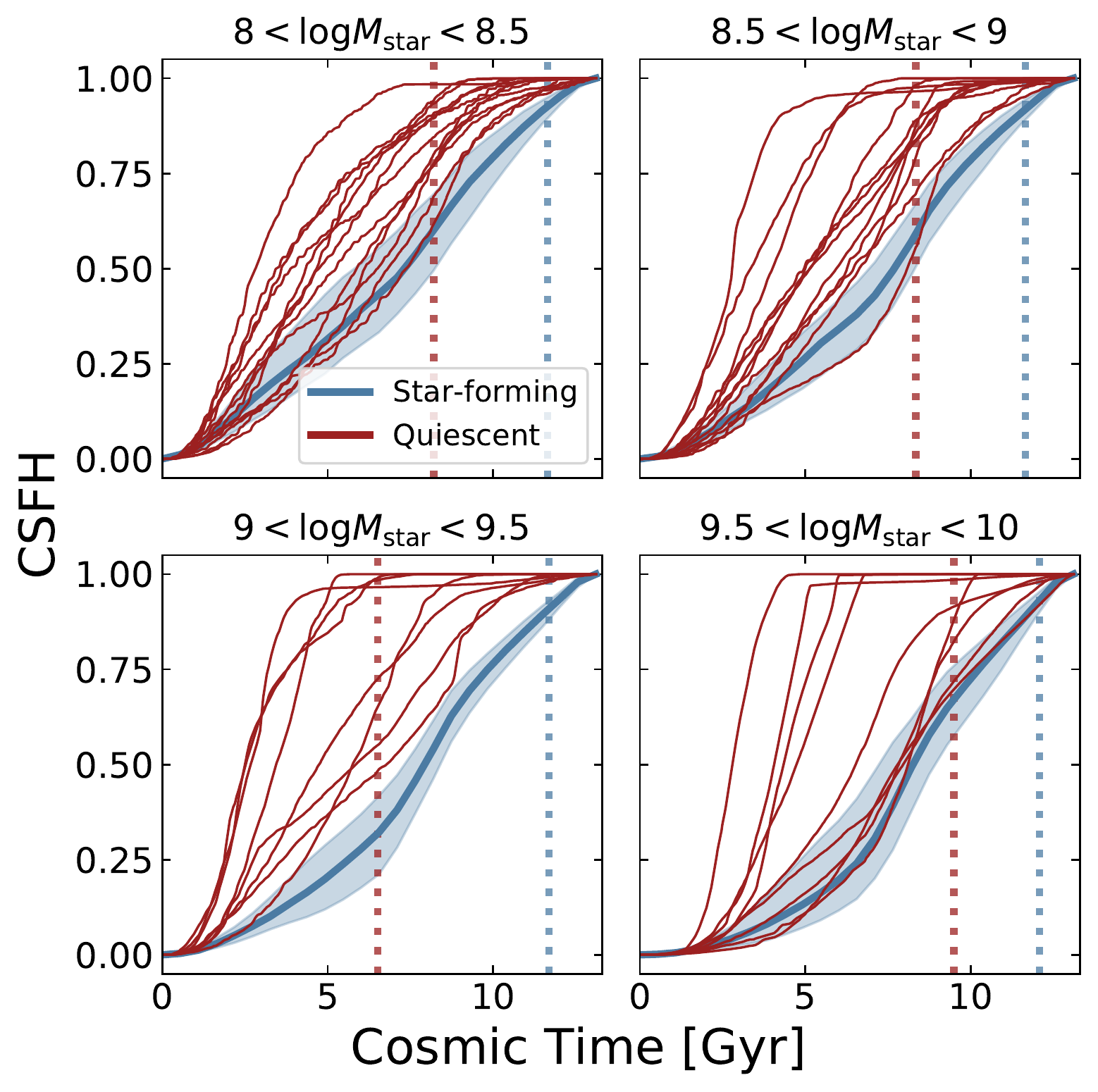}{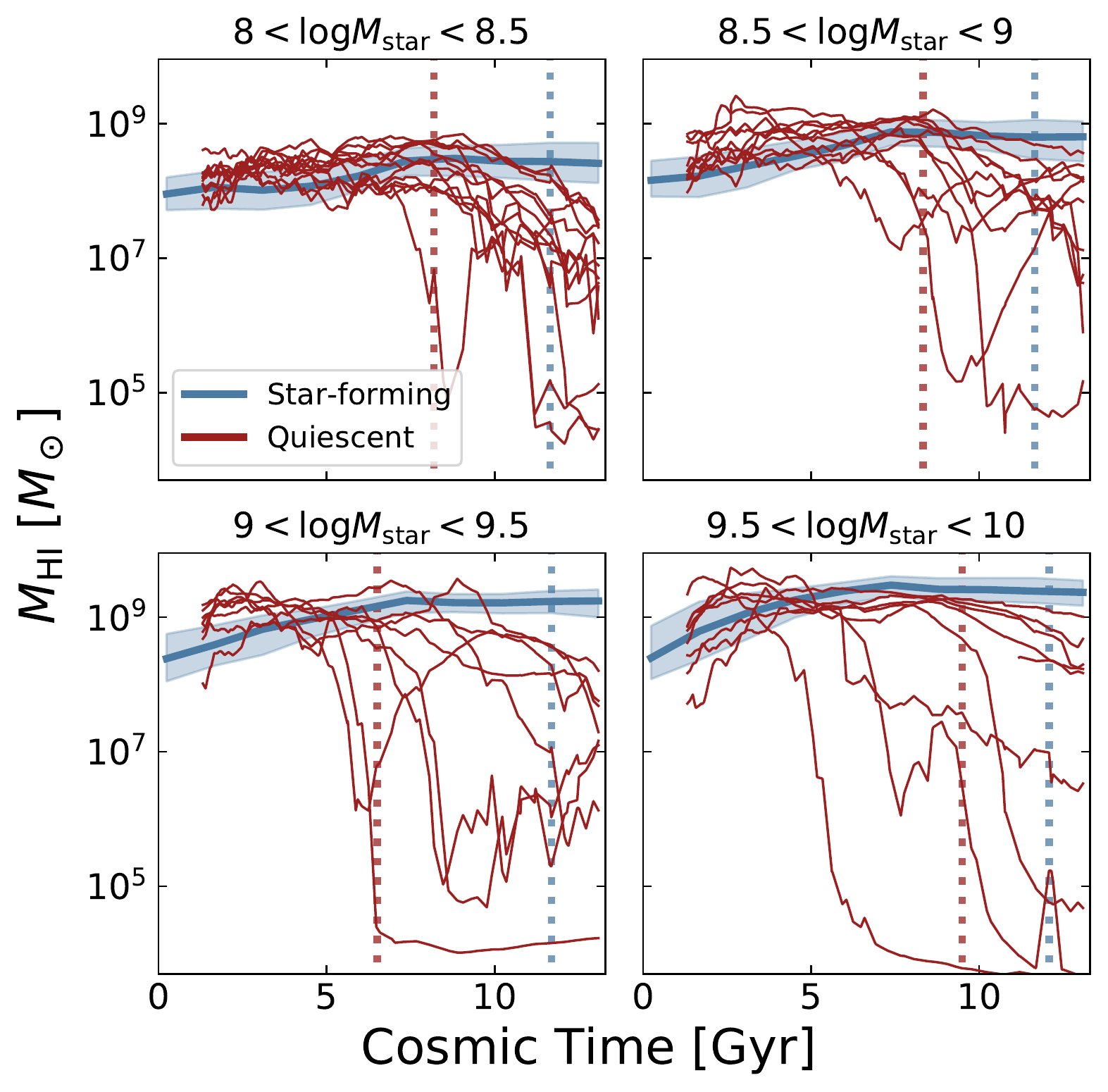}
            \caption{Cumulative star formation and \HI gas mass growth for isolated dwarfs, binned by $z=0.05$ stellar mass. Individual growth tracks are shown for quiescent dwarfs (red lines), while the median growth track for star-forming dwarfs (blue line) is shown with interquartile ranges. We show the median value of $t_{\rm 90}$ for quiescent (red dashed) and star-forming (blue dashed) dwarfs.
            {\it Left}: Cumulative star formation history over cosmic time for isolated dwarfs. In all panels, quiescent dwarfs grow stars more rapidly than star-forming dwarfs, and quench earlier. More massive dwarfs tend to quench at earlier times than less massive dwarfs, but all generally quench by redshift $z = 0.5$. {\it Right}: Total mass of cool, star-forming \HI gas as a function of cosmic time. On average, star forming dwarfs grow slowly in \HI mass over time. The majority of quiescent dwarfs instead rapidly lose \HI gas around the time of $t_{\rm 90}$, and remain devoid of \HI. Six quiescent dwarfs regain small amounts of \HI after the initial episode, reigniting low levels of star formation.}
            \label{galaxy-growth}
        \end{figure*}
        
        \begin{figure*}
            \plottwo{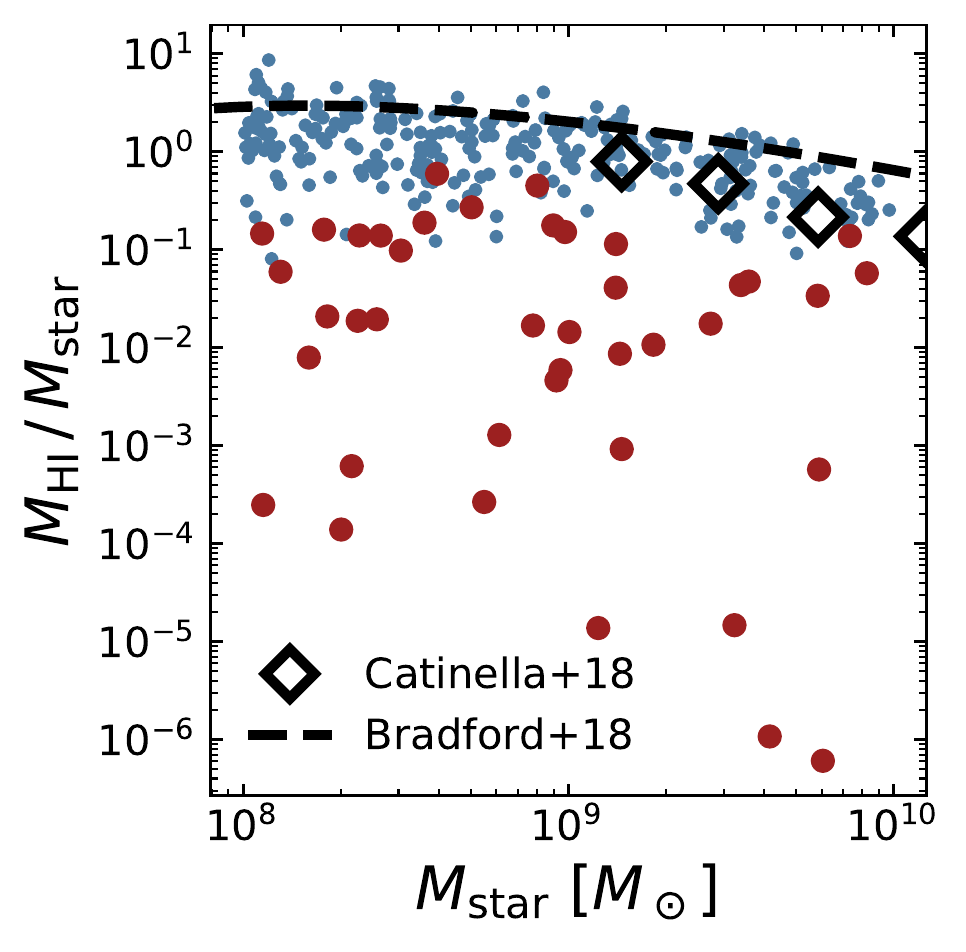}{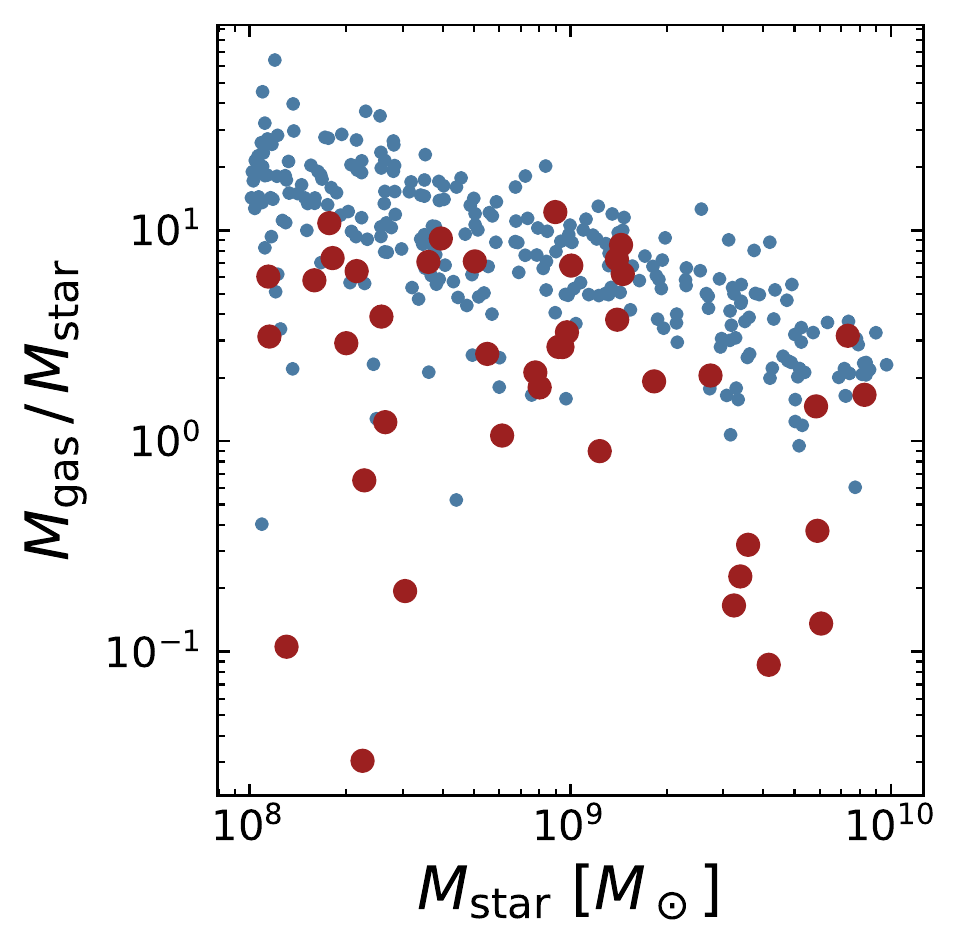}
            \caption{Fraction of gas to stars versus stellar mass within quiescent dwarfs (red circles) and star-forming dwarfs (blue dots). {\it Left}: The fraction of \HI gas to stellar mass, with observations from \citet[][black diamonds]{Catinella2018} and \citet[][black dashed]{Bradford2018} over-plotted. Quiescent dwarfs contain orders of magnitude less \HI gas at fixed stellar mass relative to star-forming dwarfs. {\it Right}: The fraction of total gas content to stellar mass. Quiescent dwarfs contain less cool, star-forming gas, and generally less gas overall relative to star-forming galaxies.}
            \label{gas-fraction}
        \end{figure*}

        With the knowledge that the presence of MBHs correlates with quiescence, we next examine the host galaxy properties in \Rom{}. In order to understand how dwarf galaxies stop forming stars, it is important to understand the gas content of these dwarf galaxies and how they evolve over cosmic time. We explore the growth histories of our dwarfs via their cumulative star formation and \HI gas mass histories. We also calculate the \HI and total gas fractions for star-forming and quiescent dwarfs.
        
        The left side of Figure \ref{galaxy-growth} shows the individual cumulative star formation histories (CSFHs) across cosmic time for quiescent galaxies along with the median for star-forming galaxies. Starting from $z=0.05$, we trace back the formation times and masses of all star particles. We divide our galaxies into bins of stellar mass at $z=0.05$. The shaded regions mark the interquartile ranges for star-forming galaxies. In each mass bin, we mark the median value of $t_{\rm 90}$, the cosmic time at which galaxies accumulate $90\%$ of their maximum stellar mass.
        
        As expected, quiescent dwarfs of all masses tend to follow different stellar growth tracks from star-forming dwarfs. Quiescent dwarfs tend to build up their stars at early times then stop forming new stars, with median $t_{\rm 90}$ approximately $5-8$ Gyr ago, or redshift $z\sim0.5-1$. Star-forming galaxies instead undergo steady formation of stars until the present day, building up their stellar component more slowly relative to quiescent galaxies. It is worth noting that not all quiescent dwarfs follow the schema of early stellar build-up then quenching -- a handful build up stars more slowly and stop at late times, following a stellar evolution closer to that of star-forming galaxies.
        
        The right side of Figure \ref{galaxy-growth} instead illustrates the history of $M_{\rm HI}$ across cosmic time for star-forming and quiescent galaxies, in the same stellar mass bins. Starting from $z=0.05$, we trace back the merger tree of each target halo and identify the main progenitor within each simulation snapshot, calculating the \HI gas mass along the way. We show the median and interquartile ranges for star-forming galaxies and plot individual evolutionary tracks for the quiescent galaxies. We also mark the median values of $t_{\rm 90}$ in each stellar mass bin for both quiescent and star-forming galaxies.
        
        On average, quiescent galaxies tend to lose \HI gas mass around the time of $t_{\rm 90}$ and maintain low $M_{\rm HI}$ afterward. Notably, six of the $42$ isolated dwarfs that are quiescent at $z=0.05$ underwent a single replenishment of their \HI gas supply, temporarily rejuvenating low levels of star formation in the recent past.
        
        For each isolated dwarf, we next measure the neutral hydrogen gas fraction, $M_{\rm HI}\,/\,M_{\rm star}$, where $M_{\rm HI}$ is the mass of gas held in an \HI component; as well as the total gas fraction, $M_{\rm gas}\,/\,M_{\rm star}$, where $M_{\rm gas}$ is the mass of all gas within the halo. Neutral hydrogen is primarily located within cool gas ($T < 2\times10^{4} K$) and is a good tracer for the supply of star-forming gas within a galaxy.
        
        The left panel of Figure \ref{gas-fraction} shows the \HI gas fraction for isolated dwarf galaxies. We distinguish between star-forming and quiescent galaxies, and compare with observed \HI gas fractions from the xGASS survey \citep{Catinella2018} and from combined ALFALFA $70\%$ observations \citep{Giovanelli2005,Haynes2011} with follow-up Arecibo observations \citep{Bradford2018}. The sample from \citet{Catinella2018} is gas-fraction limited, with galaxies selected by stellar masses between $10^{9} < M_{\rm star}/M_\odot < 10^{11.5}$ out to $z < 0.05$. The sample from \citet{Bradford2018} includes unresolved \HI observations from ALFALFA $70\%$ along with deeper Arecibo observations of galaxies with stellar masses $10^{7} < M_{\rm star}/M_\odot < 10^{11.5}$ out to $z < 0.055$. The right panel of Figure \ref{gas-fraction} shows the total gas fraction for the same subsamples of dwarfs.
        
        Quiescent dwarfs at all stellar masses exhibit a high degree of total gas depletion as well as $\sim 1$ dex lower abundance of cool, star-forming gas. Star-forming dwarfs instead follow the observed relations between \HI gas fraction and $M_{\rm star}$. Depletion of \HI gas is indicative of interstellar medium (ISM) heating and dispersal, while depletion of total gas indicates the ISM and circumgalactic medium are being removed from the halo entirely. The depleted \HI gas fractions among quiescent dwarfs is consistent with observations which find that low specific star formation rates correlate with \HI gas depletion among a sample of $25,000$ mass-selected, low-redshift galaxies from SDSS with \HI detections from ALFALFA \citep{Brown2015}. Quiescent galaxies exhibit diminished total gas fraction, though to a lesser degree than is found among \HI gas fractions. In other words, nearly all quiescent dwarfs show evidence of ISM heating, but not all quiescent dwarfs have had significant amounts of gas removed entirely from the halo. It is worth noting that the agreement of star-forming galaxies with observed relations follows from the simulation parameter tuning process \citep{Tremmel2017}, where the specific parameters controlling star formation were tuned to match \HI gas fractions from SHIELD + ALFALFA \citep{Cannon2011,Haynes2011}. However, the relationship of quiescence with gas depletion is a prediction of the simulation.
        
        Overall, these trends give insight into how our dwarf galaxies stop forming stars. Gas that may otherwise form stars is instead kept hot and/or completely driven out, while replenishing inflows of cold gas from the circumgalactic and intergalactic medium are suppressed. We examine the flow of gas more explicitly for a few cases in Section \ref{case-studies}. This combination of gas ``starvation'' \citep{Peng2015} and gas depletion is reminiscent of the mechanisms invoked in recent studies to explain quenching in local low-mass galaxies. \citet{Trussler2020} study the chemical properties of optically selected local galaxies with stellar masses $M_{\rm star} > 10^{9} M_\odot$, showing that passive star formation in their sample is likely driven by a combination of gas starvation and {\it ejection} via outflows. \citet{Bluck2020} similarly train a neural network on spatially resolved spectroscopic observations of $3500$ local $M_{\rm star} > 10^{9} M_\odot$ galaxies from MaNGA \citep{Aguado2019}, showing that global properties of central galaxies (e.g., central velocity dispersion) are more powerful predictors of spaxel-wise quiescence than even spaxel-wise properties (e.g., stellar mass surface density per IFU spaxel). They find that galaxy-wide gas starvation alongside MBH feedback are the primary source of quenching among central galaxies.

    \subsection{Interactions rarely lead to quenching}
    \label{sec:interactions}
    
        \begin{figure}
            \plotone{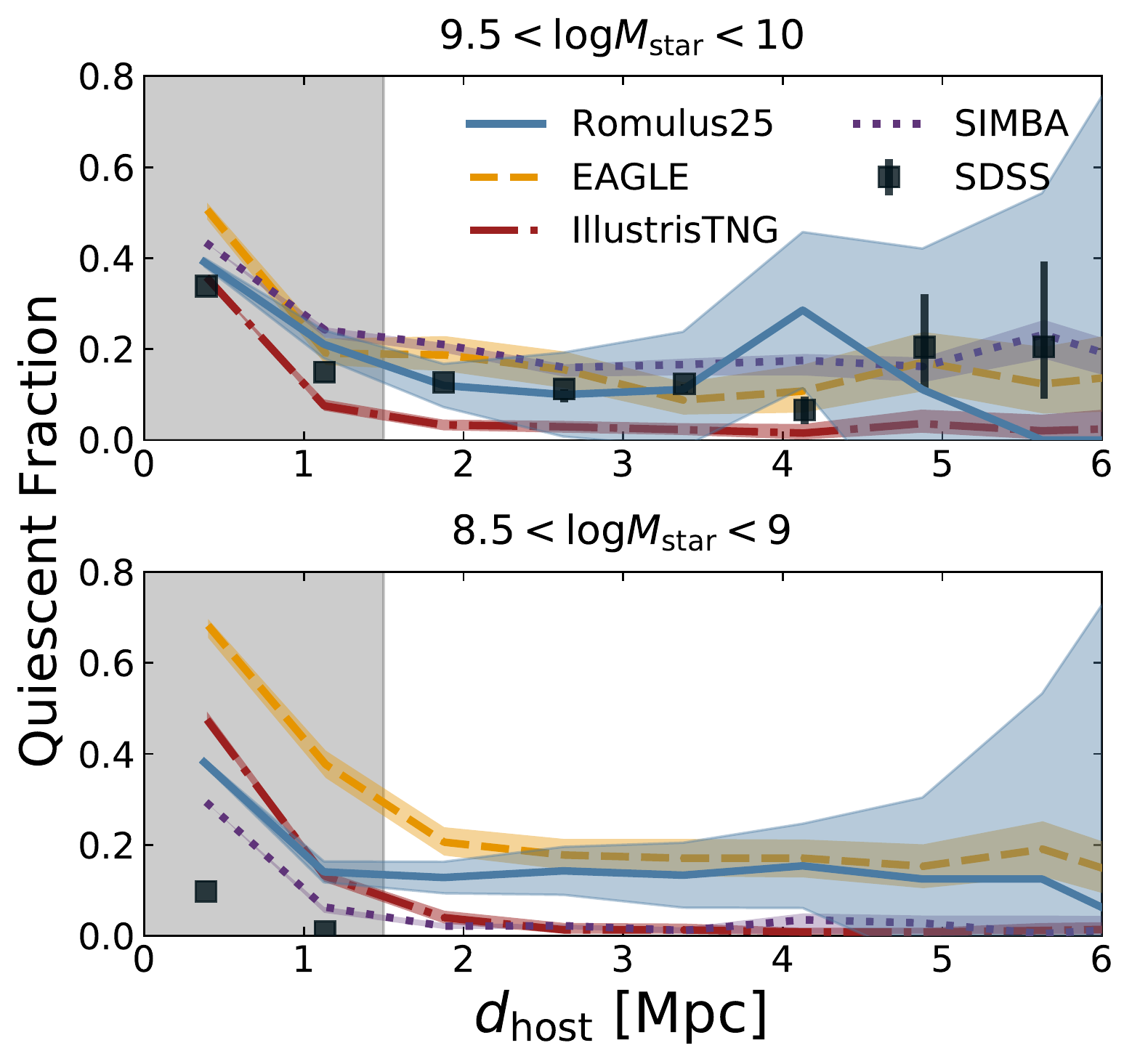}
            \caption{Quiescent fraction across $25$ sight-lines as a function of distance from the nearest massive neighbor. Legend is as in Figure \ref{qfrac-comparison}. For comparison with \citet{Dickey2021}, we distinguish between high-mass dwarfs (top panel, $10^{9.5} < M_{\rm star}/M_\odot < 10^{10}$) and low-mass dwarfs (bottom panel, $10^{8.5} < M_{\rm star}/M_\odot < 10^{9}$). The black shaded region indicates $d_{\rm host} < 1.5$ Mpc, the threshold for galaxy isolation. \Rom{} dwarfs (isolated and non-isolated) exhibit the same trends as most other simulations, showing quiescent fractions in agreement with observations for high-mass dwarfs, but predicting too many quiescent low-mass dwarfs. At low masses, all simulations produce too many quiescent dwarfs even within group environments with $d_{\rm host} < 1.5$ Mpc.}
            \label{qfrac-dhost}
        \end{figure}
        
        \begin{figure}
            \plotone{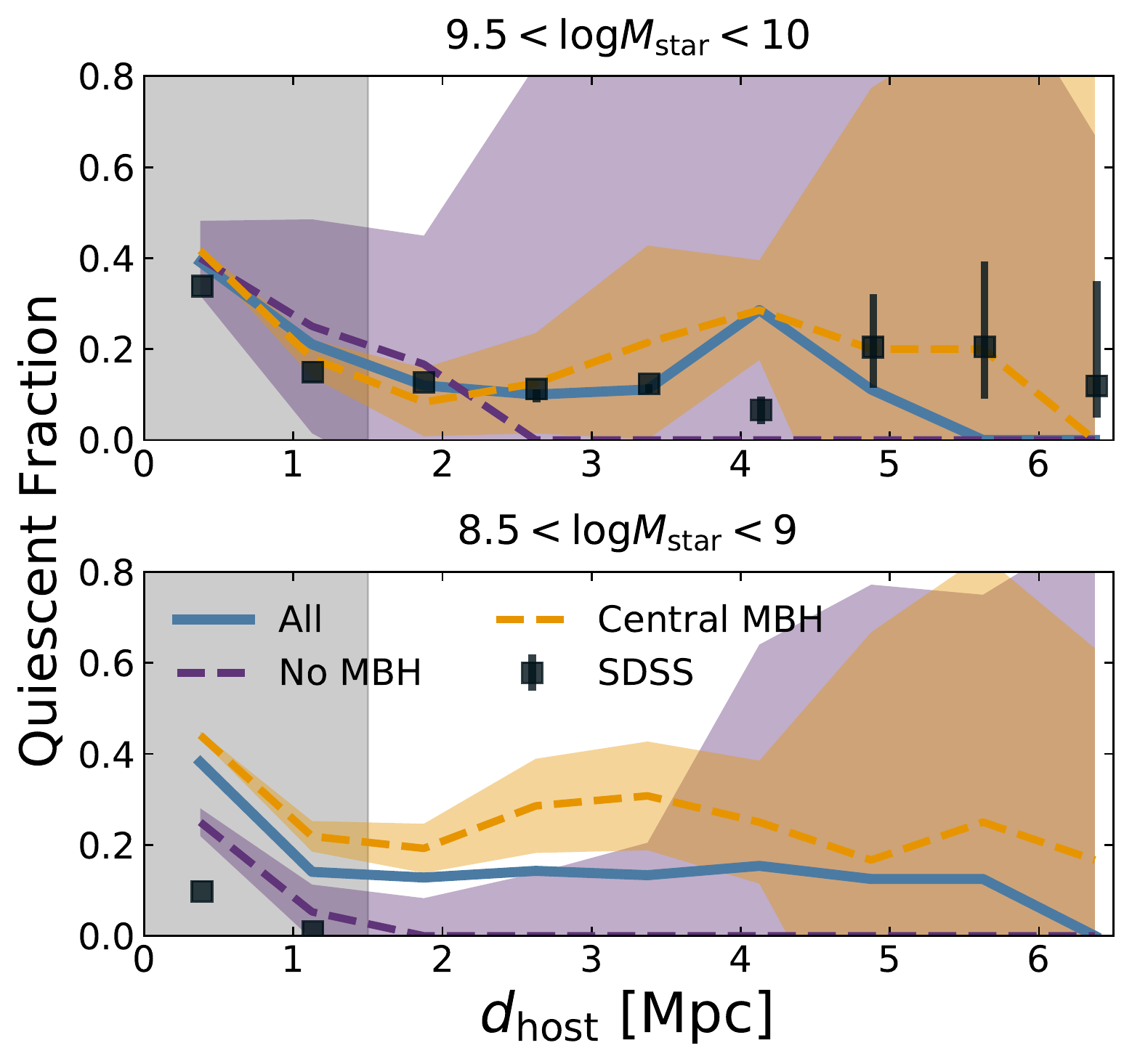}
            \caption{Quiescent fraction across $25$ sight-lines in \Rom{} as a function of distance from the nearest massive neighbor, distinguishing between all dwarfs (blue), those with central MBHs (yellow), and those with no MBH (purple). As in Figure \ref{qfrac-dhost}, we distinguish between high-mass dwarfs (top panel, $10^{9.5} < M_{\rm star}/M_\odot < 10^{10}$) and low-mass dwarfs (bottom panel, $10^{8.5} < M_{\rm star}/M_\odot < 10^{9}$). Uncertainties are omitted on the ``All'' line for clarity. The grey shaded region marks $d_{\rm host} < 1.5$ Mpc, the threshold for galaxy isolation. High-mass dwarfs with central MBHs are more likely to be quenched in the field relative to those without MBHs. Low-mass \Rom{} dwarfs with central MBHs are more likely to be quenched in both the field and in groups, relative to those without MBHs. Dwarfs without MBHs in both mass ranges show closer agreement to observations of the quiescent fraction \citep[][black squares]{Dickey2021}.}
            \label{qfrac-dhost-bhs}
        \end{figure}

        Before we can fully assess the impact of internal, MBH-driven processes on star formation, it is important to rule out the possibility of interactions driving the majority of quiescence. By selecting galaxies in isolation, we have mitigated (but not eliminated) the effects of environmental quenching on our dwarfs. We first ask how quenching compares among dwarfs in isolation relative to those with neighbors. 
        
        In Figure \ref{qfrac-dhost}, we show the quiescent fraction at $z=0.05$ as a function of distance to the nearest host, $d_{\rm host}$. As in \citet{Dickey2021}, we separate galaxies into two stellar mass bins: just below the \citet{Geha2012} quenching threshold $\left(10^{8.5} < M_{\rm star}/M_\odot < 10^{9}\right)$, and just above $\left(10^{9.5} < M_{\rm star}/M_\odot < 10^{10} \right)$. As before, we show the median across all $25$ sight-lines and compare with results from other cosmological simulations as well as SDSS observations.
        
        Similar to other simulations, both quiescent isolated and quiescent non-isolated galaxies in \Rom{} are produced in excess relative to SDSS observations. In the higher mass bin, \Rom{}, {\sc EAGLE}, and {\sc SIMBA} produce quiescent fractions consistent with SDSS for $d_{\rm host} > 1.5$ Mpc, though most simulations slightly overproduce quiescent galaxies for $d_{\rm host} < 1.5$ Mpc. In the lower mass bin, however, all simulations produce too many quiescent non-isolated galaxies, to varying degrees. In particular, \Rom{} yields non-isolated quiescent fractions between $20-30\%$, substantially higher than the $0-10\%$ found in SDSS. It is possible that the mechanism responsible for quenching isolated galaxies in the simulations is also impacting quenching in dense environments, and may explain the over-production of low-mass quiescent dwarfs in dense environments.
        
        We may also ask how quiescence varies for dwarfs with and without MBHs as a function of environment. Figure \ref{qfrac-dhost-bhs} shows the quiescent fraction as a function of distance to the nearest massive neighbor, distinguishing between dwarfs with central MBHs and those without MBHs. As in Figure \ref{qfrac-dhost}, we split high- and low-mass dwarfs. 
        
        High-mass dwarfs with central MBHs tend to follow observed quiescent fractions in both the group and field environments. High-mass dwarfs without MBHs follow observed quiescent fractions in group environments, though drop more rapidly with $d_{\rm host}$ than observations. On the other hand, low-mass dwarfs with central MBHs tend to over-predict the quiescent fraction in both the group and field environments. As seen earlier, dwarfs with central MBHs drive much of the discrepancy between \Rom{} and observed quiescent fractions, and indeed low-mass dwarfs without MBHs show much closer agreement with observations.
        
        It is important to note that there are cases of splashback \citep[e.g,][]{Tinker2017} in our sample, where a galaxy that is currently isolated was within close proximity of a larger halo in the past. By tracing back each target halo's interaction history, we identify times where the target halo first fell within the radius of a more massive halo, and compare these to the quenching times, $t_{\rm 90}$.
        
        Overall, we find that splashback is unlikely to play a major role in quenching our dwarf galaxy sample. Of the $328$ galaxies in our sample, $60$ $\left(20\%\right)$ have undergone splashback in the past. Approximately $17\%$ of splashback galaxies are quiescent, only slightly higher than the $12\%$ of non-splashback galaxies that are quiescent. Conversely, $24\%$ of quiescent galaxies are splashback, only slightly higher than the $17\%$ of star-forming galaxies that are splashback. The number of splashback galaxies is dependent on stellar mass, where dwarfs below $M_{\rm star}/M_\odot < 10^{9}$ make up $77\%$ of splashback galaxies. Still, these lower mass dwarfs show similar quiescent fractions between those that are splashback $(17\%)$ and non-splashback $(11\%)$.
        
        These splashback results are in line with findings from recent zoom cosmological simulations, which have found that low-mass satellites $\left(M_{\rm star}/M_\odot \lesssim 10^{8}\right)$ continue forming stars for $2-5$ Gyr after infall into the host halo \citep{Akins2021}, while higher mass satellites $\left(M_{\rm star}/M_\odot \gtrsim 10^{8}\right)$ typically never fully quench \citep{Akins2021,Samuel2022}. While a small number of our isolated dwarfs likely quench due to splashback, the majority do not.
        
        Finally, we determine whether mergers are the primary mode by which quiescence was catalyzed. Mergers are thought to dissipate gas angular momentum, channeling gas and driving quenching via a starburst \citep[e.g,][]{Wild2009,Wild2016} or AGN activity \citep[e.g,][]{Hopkins2006b,Hopkins2008a}. We climb each target halo's merger tree backward in time in search of cases where the next, earlier snapshot contains more than one progenitor. We restrict our search to mergers with halo mass ratios $>1:10$ at the snapshot time. Identifying mergers in this way misses the effects of minor progenitors that become stripped over time, but captures the effects of major mergers. The merger time, $t_{\rm merge}$, is then the cosmic time of the snapshot containing only the descendent of the merger. As in \citet{McAlpine2020}, we define the dynamical time in terms of the cosmic critical density, $\rho_{\rm c}$,
        \begin{align}
            t_{\rm dyn} = \left(\frac{3\pi}{32 G \left(200 \rho_{\rm c}\right)}\right)^{1/2},
        \end{align}
        and identify dwarfs that underwent mergers within one dynamical time of the quenching time, $\left|t_{\rm merge} - t_{\rm 90}\right| < t_{\rm dyn}$.
        
        Mergers are unlikely to be the primary mechanism by which quenching is initiated. Only $11\%$ of quiescent dwarfs have undergone significant mergers within one dynamical time of $t_{\rm 90}$, similar to the $10\%$ of star-forming dwarfs. Relaxing our criteria, we find that $21\%$ of quiescent dwarfs have undergone a merger within two dynamical times of quenching, similar to the $18\%$ of star-forming dwarfs. Evidently, mergers are not a hard requirement to initiate quenching of isolated dwarf galaxies.
        
    \subsection{Internal processes can quench isolated dwarfs} \label{feedback}

        \begin{figure*}
            \plotone{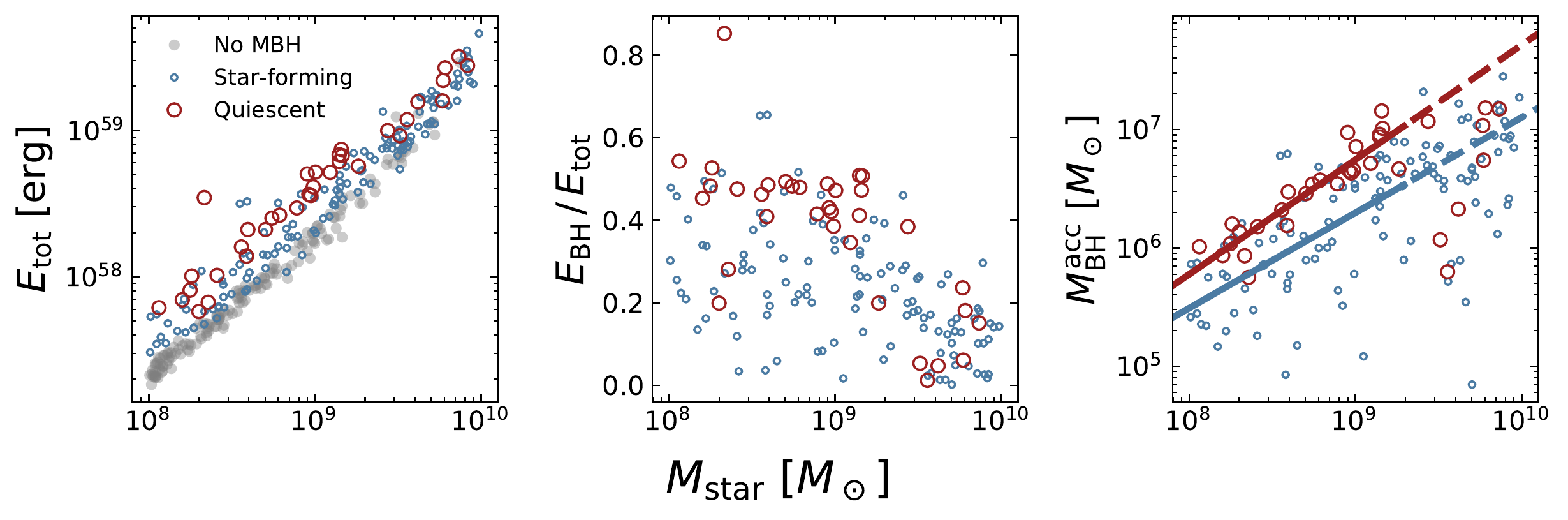}
            \caption{{\it Left}: Total energy injection from MBHs and SNe as a function of stellar mass for star-forming dwarfs with central MBHs (blue unfilled dots), quiescent dwarfs with central MBHs (red unfilled circles), and dwarfs entirely without MBHs (grey filled circles).  {\it Middle}: Fraction of energy injection contributed by MBHs as a function of stellar mass for dwarfs with central MBHs. Quiescent dwarfs below $M_{\rm star}/ M_\odot < 2\times10^{9}$ tend to have higher amounts of total energy injection at fixed stellar mass, and more of this energy injection tends to come from MBHs. Dwarfs above $M_{\rm star}/ M_\odot > 2\times10^{9}$ have similar amounts of total energy injection, in-line with simulation results from \citet{Pontzen2017}. Only a small fraction of dwarfs are entirely dominated by energy injection from the MBH. {\it Right}: $M_{\rm BH}^{\rm acc} - M_{\rm star}$ relation for quiescent and star-forming dwarfs with central MBHs. We fit the relation below $M_{\rm star}/ M_\odot < 2\times10^{9}$ for quiescent galaxies (red solid) and star-forming galaxies (blue solid), showing the extrapolated fits with dashed lines.}
            \label{etot-mstar}
        \end{figure*}
        
        Turning now to the precise effects of feedback, we calculate the integrated energy injection from both MBHs and Type II supernovae (SNe II) into the surrounding ISM. There is precedent for exploring the underlying energetics of AGN in relation to quenching. Using high resolution cosmological simulations of a $M_{\rm vir} = 10^{12} M_\odot$ halo, \citet{Pontzen2017} showed that galaxy quenching is related to a greater contribution from MBH activity to the total energy injected. Similarly, \citet{Dashyan2018} use analytic models of AGN energetics to show that large regions of the AGN efficiency -- lifetime parameter space in dwarf galaxies are more favorable to AGN-driven gas removal versus SNe-driven removal.
        
        \citet{Dashyan2018} use analytic models to show that feedback from AGN may be even more successful than SNe in driving out gas from galaxies below a critical halo mass threshold.
        
        Within \Rom{}, the integrated energy injected into the surrounding gas by SNe II is directly related to the number of SNe II that have detonated within the galaxy, $N_{\rm SN}$. Although other types of SNe are implemented, we ignore the effects of Type Ia SNe since they do not drive large-scale blastwave feedback in \Rom{} \citep{Stinson2006}. Since the timing of the SNe II depends on the mass of the stellar progenitor, integrating over the IMF provides the number of stars that will explode within a given star particle, where only stars between $8 - 40 M_\odot$ are allowed to detonate as SNe II \citep{Stinson2006}. Each SNe II event injects the same amount of energy into the surrounding gas, ${\rm d}E_{\rm SN} = 0.75 \times 10^{51}\,{\rm erg}$ \citep{Tremmel2017}. The integrated SNe energy injection can then be written:
        \begin{align}
            E_{\rm SN} = N_{\rm SN} {\rm d}E_{\rm SN}.
        \end{align}
        
        The energy injected by the MBH into the surrounding ISM is directly related to the accretion rate, $\dot{M}$, via the radiative efficiency, $\epsilon_{\rm r}=0.1$, and the gas coupling efficiency, $\epsilon_{\rm f} = 0.02$ \citep{Tremmel2017}. The integrated MBH energy injection can then be written:
        \begin{align}
            E_{\rm BH} = \epsilon_{\rm r} \epsilon_{\rm f} \dot{M} c^2 {\rm d}t,
        \end{align}
        where the accretion is assumed to be constant throughout one MBH simulation timestep, ${\rm d}t$.
        
        The left panel of Figure \ref{etot-mstar} shows the total energy injected into each dwarf's ISM, $E_{\rm tot} = E_{\rm BH} + E_{\rm SN}$, as a function of stellar mass, where we distinguish between quiescent and star-forming dwarfs with central MBHs, and dwarfs entirely without MBHs. 
        Since the number of SNe II events scales directly with number of star particles, $E_{\rm SN}$ is closely tied to the stellar mass of the galaxy. Hence at fixed stellar mass, the difference between dwarfs without MBHs and those with central MBHs is largely due to energy injection from the MBH. 
        
        The MBH energy injection is seen in more detail in the center panel of Figure \ref{etot-mstar}, where we instead show the fractional contributions of MBHs to the total energy injected, $\left(E_{\rm BH}\right)\,/\,\left(E_{\rm BH} + E_{\rm SN}\right)$. Here, we only include dwarfs with central MBHs. Among dwarfs with $M_{\rm star}/ M_\odot < 2\times10^{9}$, the fraction of the energy budget attributable to MBHs is higher in quiescent dwarfs than in star-forming dwarfs. Above $M_{\rm star}/ M_\odot > 2 \times 10^{9}$, dwarfs shift to strongly SNe-dominated, with little difference in energy injection between quiescent and star-forming galaxies. 
        
        The latter behavior is slightly different than results from \citet{Pontzen2017}, who study cosmological simulations of a $M_{\rm vir} = 10^{12} M_\odot$ halo at $z=2$. They find that the total energy injection from feedback is comparable between MBHs+SNe runs and SNe-only runs of their halo, despite quenching only occurring in the MBH+SNe runs. However, they find that $E_{\rm BH} > E_{\rm SN}$ in cases where the halo quenches. In contrast, our lower mass dwarfs with MBHs have slightly enhanced $E_{\rm tot}$, and only $15\%$ of all quiescent dwarfs and $3\%$ of all star-forming dwarfs exhibit $E_{\rm BH} > E_{\rm SN}$. Evidently, MBHs only need to contribute a significant amount of energy in order to quench, and do not necessarily need to dominate the energy budget. The importance of MBH energy injection in suppressing star formation is in agreement with results from \citet{Piotrowska2022}, who study the quenching mechanisms in $M_{\rm star} > 10^{9} M_\odot$ central galaxies in the {\sc EAGLE}, {\sc Illustris}, and {\sc IllustrisTNG} cosmological simulations. Using a random forest classifier, they find that classification of a galaxy as star-forming or quiescent is most closely tied to the integrated accretion history of the central MBH. However, as discussed in Section \ref{quiescent-fractions}, the other simulations considered in this work do not seed MBHs in galaxies below $M_{\rm star} < 10^{9} M_\odot$. \citep{Dickey2021} find that the over-production of low-mass quenched dwarfs in these other simulations is instead due to either environmental effects, feedback from SNe, or outflows from nearby massive galaxies.
        
        As discussed above, the energy injection from SNe II and MBHs are closely related to the stellar mass and accreted MBH mass, respectively. The right panel of Figure \ref{etot-mstar} shows the $M_{\rm BH}^{\rm acc} - M_{\rm star}$ relation for our isolated dwarf sample with central MBHs. Linear fits to the $M_{\rm BH}^{\rm acc} - M_{\rm star}$ relation for quiescent and star-forming galaxies below $M_{\rm star}/M_\odot < 2\times10^{9}$ indicate that quiescent dwarfs host MBHs $3\times$ more massive than the MBHs in corresponding star-forming dwarfs. This phenomenon is directly in-line with results from \citet{Sharma2020}, which found that over-massive MBHs are related to star-formation suppression and gas depletion in \Rom{} dwarf galaxies. However, this behavior breaks down for galaxies above $M_{\rm star}/M_\odot > 2\times10^{9}$, where three quiescent galaxies in fact host under-massive MBHs. The quenching mechanism for dwarfs above and below $M_{\rm star}/M_\odot = 2\times10^{9}$ clearly differ, and will require closer inspection.
    
    \subsection{Case Studies} \label{case-studies}
    
        \begin{figure*}
            \plotone{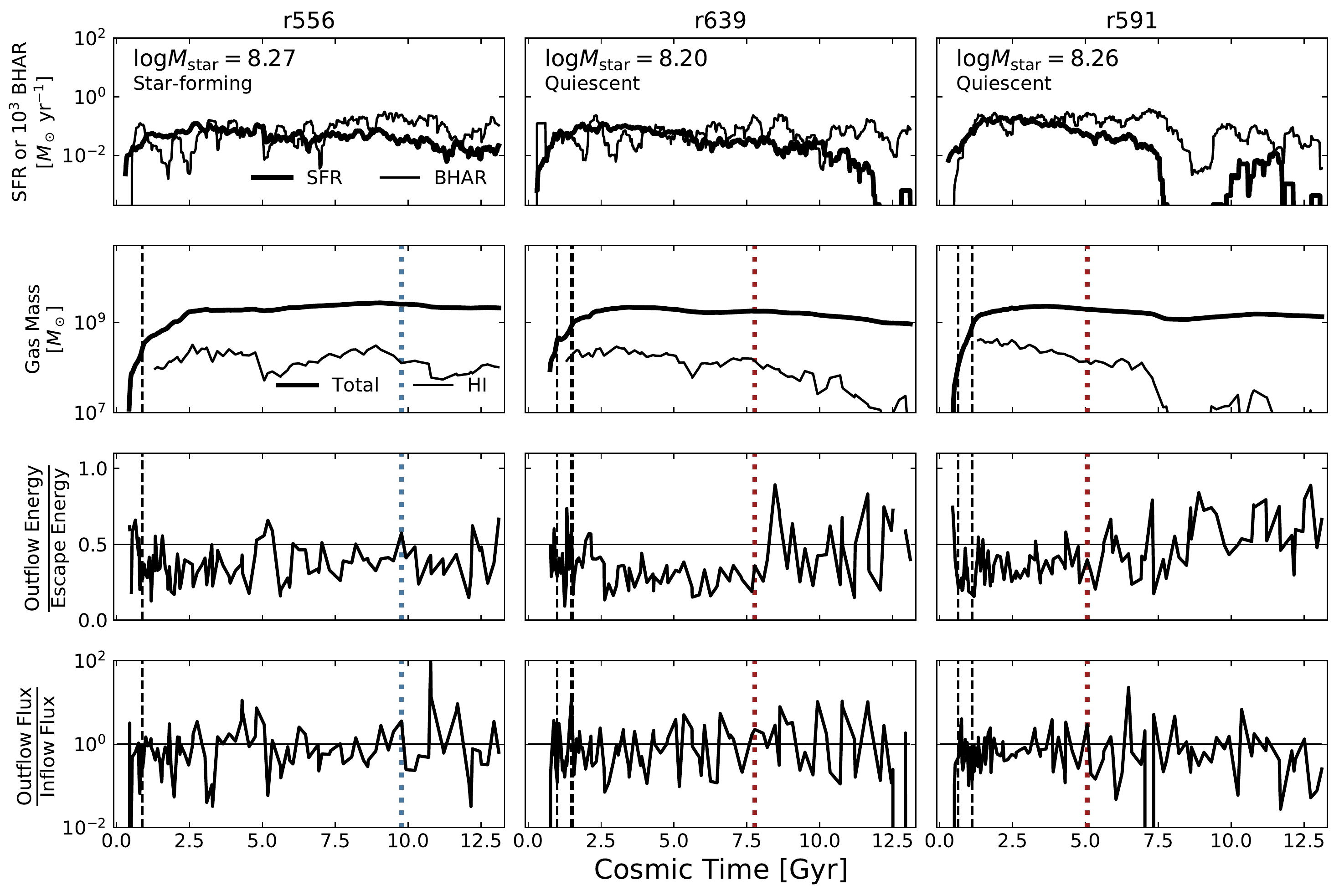}
            \caption{History of gas kinematics, star formation, and MBH accretion for three low-mass dwarf galaxies between $10^{8} < M_{\rm star}/M_\odot < 10^{8.3}$, with one star-forming dwarf (left) and two quiescent dwarfs (center+right). We trace the properties of the most massive MBH within $2$ kpc of the halo center. Each column is labelled with the internal halo number from \Rom{}. All panels except the top row include vertical lines marking $t_{\rm 90}$ (blue dotted for star-forming, red dotted for quiescent), as well as vertical lines marking mergers (dashed-black, see Section \ref{sec:interactions} for details). 
            {\it 1st row}: SFR (solid thick) and $10^3 \times$ BHAR (solid thin) versus cosmic time. 
            {\it 2nd row}: Mass held within all gas particles (solid thick) and mass held within the \HI component of gas particles (solid thin). 
            {\it 3rd row}: Ratio of energy in outflowing gas at $0.1\times R_{\rm 200}$, to the energy required to escape the potential from $0.1 \times R_{\rm 200}$. For clarity, the horizontal line (solid thin) marks the $0.5$ threshold.
            {\it 4th row}: The ratio of mass flux in outflowing gas to inflowing gas at $0.1\times R_{\rm 200}$.}
            \label{case-studies-1}
        \end{figure*}
        
        \begin{figure*}
            \plotone{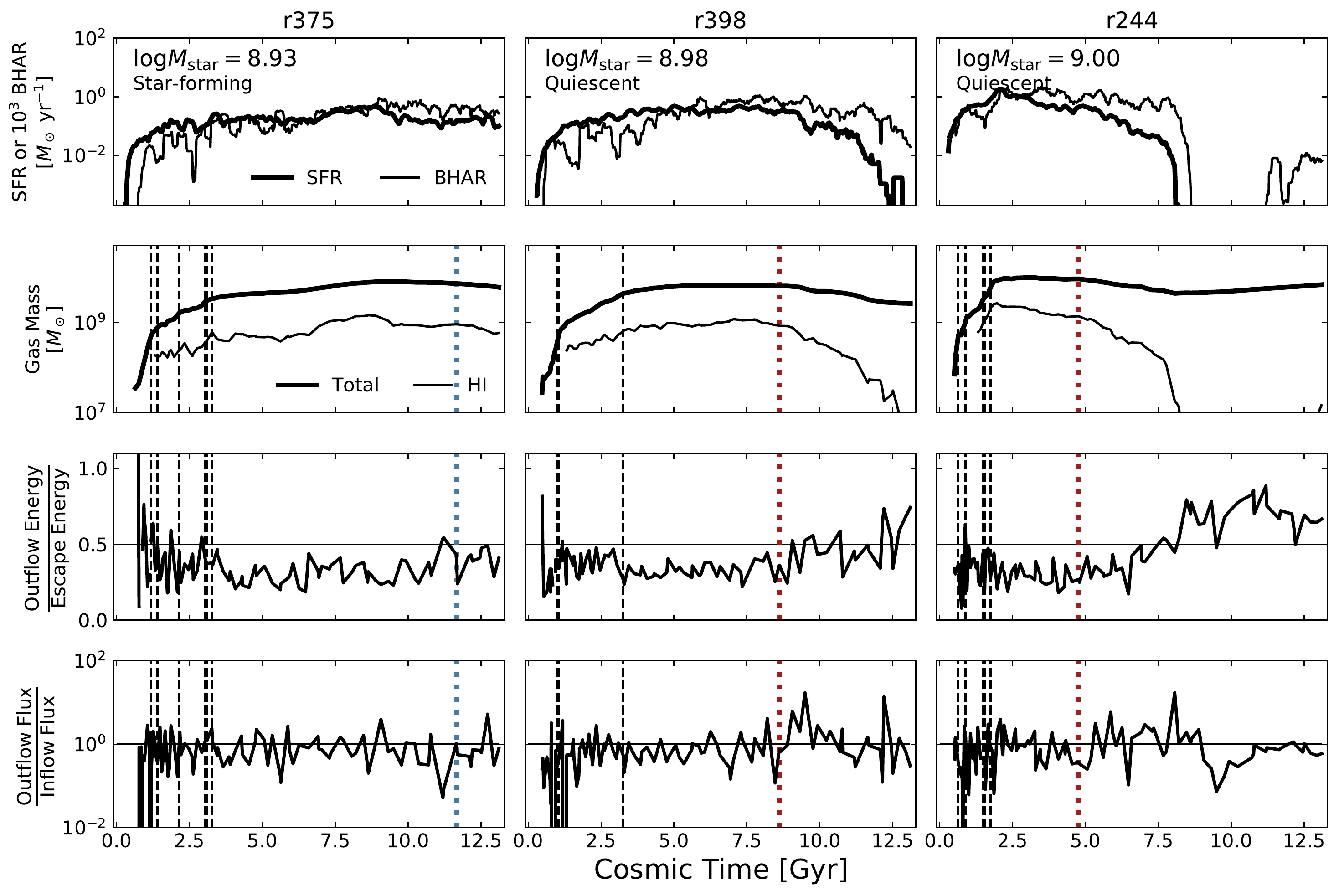}
            \caption{As in Figure \ref{case-studies-1}, but for three intermediate-mass dwarf galaxies between $10^{8.8} < M_{\rm star}/M_\odot < 10^{9.2}$. Quenching can occur both slowly as in lower mass dwarfs, or rapidly as in higher mass dwarfs.}
            \label{case-studies-2}
        \end{figure*}
        
        \begin{figure*}
            \plotone{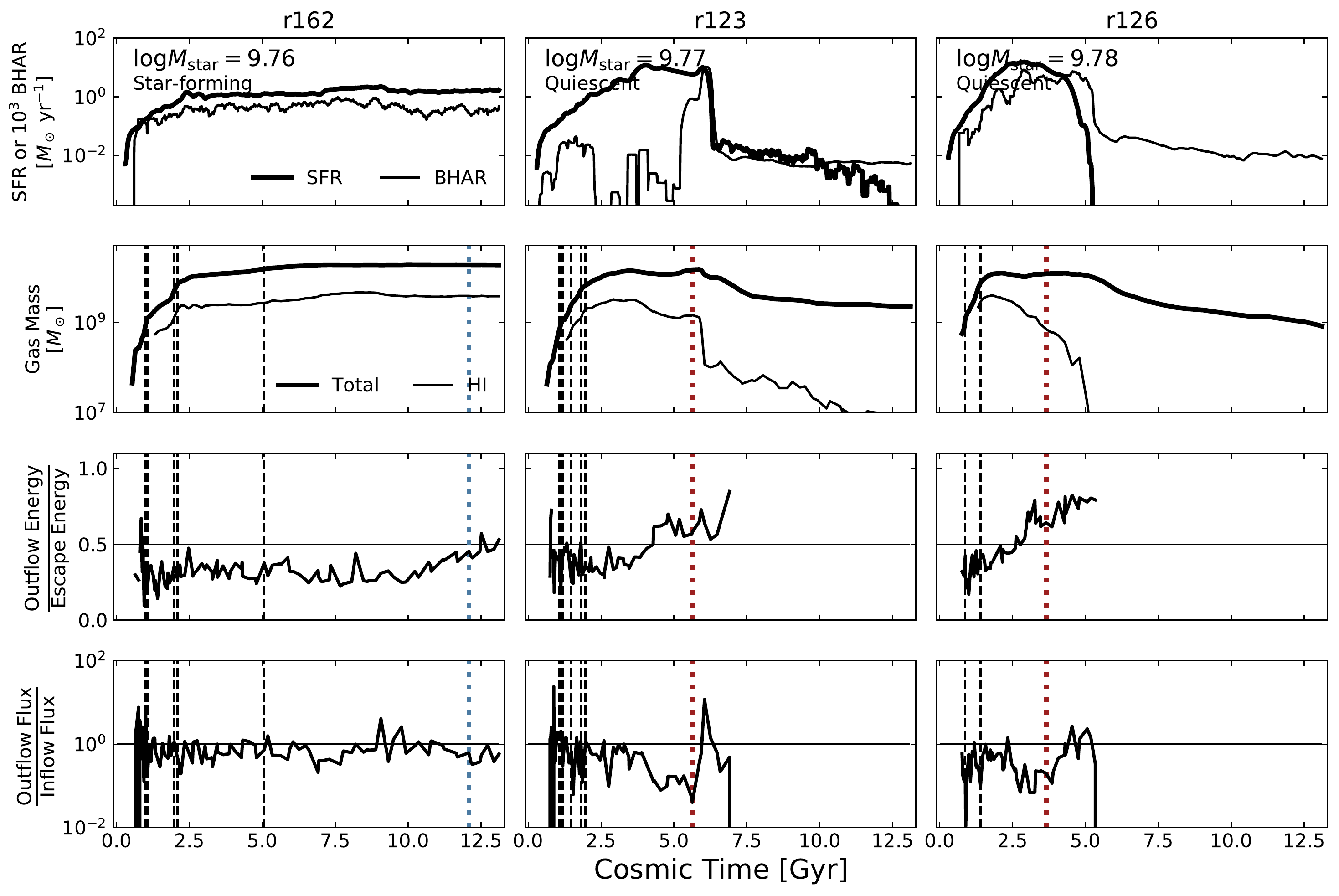}
            \caption{As in Figure \ref{case-studies-1}, but for three high-mass dwarf galaxies between $10^{9.7} < M_{\rm star}/M_\odot < 10^{10}$. Quenching occurs rapidly in a ``blow-out'' phase, in stark contrast to the slow quenching found in lower mass dwarfs. Outflows within $r123$ and $r126$ shutdown within $1$ Gyr of $t_{\rm 90}$, causing a truncation in the energy and flux panels.}
            \label{case-studies-3}
        \end{figure*}
    
        We now more closely examine the histories of gas mass, star formation, and MBH accretion for nine dwarfs as a way to understand the differences in quenching mechanisms. We divide our stellar mass range of interest into three bins, $M_{\rm star} = \left[10^{8-8.3} M_\odot, 10^{8.8 - 9.2} M_\odot, 10^{9.7 - 10} M_\odot\right]$, and randomly select one star-forming dwarf from each bin. For each star-forming galaxy, we select two quiescent galaxies with stellar masses within $0.1$ dex.
        
        For each dwarf at $z=0.05$, we trace the main halo progenitor backward in time through all available simulation snapshots. At each snapshot we record: the mass held within gas particles as well as the mass within the \HI component; the inflowing/outflowing mass flux and energy of gas flowing across a $1$ kpc thick shell of radius $0.1\times R_{\rm 200}$; and the energy required for gas to escape the gravitational potential from $0.1\times R_{\rm 200}$ out to $R_{\rm 200}$. We also identify the total SFR for the stars within the dwarf at $z=0.05$, and the accretion rate of the central MBH, which are output at finer time resolution. These quantities may yield insight into how the timing of MBH activity affects gas kinematics and ultimately star formation.
        
        Studies connecting outflows and quenching within simulations have been done in the past. \citep{Pontzen2017} and \citep{Sanchez2021} study properties of AGN outflows in massive galaxies, finding a strong connection between energetic outflows and galaxy quenching. Within the galaxy cluster environment, \citet{Tremmel2019} and \citet{Chadayammuri2021} find that large scale, AGN-driven outflows regulate gas cooling onto the brightest cluster galaxy, quenching star formation. \citet{Koudmani2021} use the \textsc{FABLE} cosmological simulations to show that dwarf galaxies with energetic, MBH-driven outflows tend to exhibit reduced gas mass fractions and contribute to dwarf quenching at high redshift.
        
        Figures \ref{case-studies-1}, \ref{case-studies-2}, and \ref{case-studies-3} show case studies for dwarfs in each of the three mass bins. In each figure, the top row traces SFR and $10^3 \times$ MBH accretion rate across time. The second row traces the total gas mass and \HI gas mass across time. The third row tracks the ratio of outflow gas energy to the escape energy \citep{Pontzen2017}, where we define the outflow energy:
        \begin{align}
            \textrm{Outflow energy} = \frac{\sum_i m_i v_{{\rm r}, i} \left(u_i + v_{{\rm r}, i}^2 / 2\right)}{\sum_i m_i v_{{\rm r}, i}},
        \end{align}
        with $m_i$ gas particle mass, $u_i$ gas internal energy, and $v_{{\rm r},i}$ gas radial velocity. We also define the escape energy:
        \begin{align}
            \textrm{Escape energy} = \Phi\left(R_{\rm 200}\right) - \Phi\left(0.1\times R_{\rm 200}\right),
        \end{align}
        for spherically averaged potential, $\Phi\left(r\right)$ at radius $r$.
        
        Dwarfs at low masses $\left(M_{\rm star} \sim 10^{8} M_\odot \right)$ quench slowly over several Gyr. The supply of star-forming gas is slowly used up or heated by both the central MBH and star formation. Although the outflow energetics indicate a quick increase in outflow energy following quenching, there are no coherent outflows that consistently drive gas outward across this radius. Further, \HI mass decreases over time while the total gas mass remains constant or slowly decreasing, suggesting that the quenching mechanism struggles to physically remove gas but can suppress cooling. Along with the fact that MBH activity continues after quenching, together these suggest that quenching for low-mass dwarfs is driven by gas heating induced by the combination of star formation and MBH activity. The quenched status is maintained by continued MBH feedback. At early times, the galaxy is able to accrete and cool gas, allowing star formation and MBH accretion to proceed until the point that heating from the MBH overwhelms ISM cooling.
        
        Dwarfs at high masses instead quench explosively (see Figure \ref{case-studies-3}), similar to what is thought to occur in more massive galaxies \citep{Hopkins2005,Ishibashi2016a}. High mass dwarfs tend to quench within a short period of time $\left(< 1 \textrm{ Gyr}\right)$, with a period of rapid gas inflow driving fast MBH accretion, which in turn heats up gas and drives it outward across $0.1\times R_{\rm 200}$. In the case of {\it r123}, both MBH accretion and star formation drop quickly but maintain low levels. Star formation in {\it r126} instead drops and stays at zero at the time of quenching, while the MBH continues to accrete at low levels. In both cases, nearly all of the gas is swept up and driven outside of $0.1\times R_{\rm 200}$. But the outflow energies indicate the gas is not able to fully escape the halo, and instead it is kept hot in the outskirts of the halo. The impact of the central MBH on the outskirts of the halo is in stark contrast with zoom simulation results from \citet{Sanchez2019,Sanchez2021}, which find that MBHs in Milky Way-mass halos can drive large outflows from the disk, but are \textit{not} able to drive changes in gas content in the circumgalactic medium. They find that MBH-induced outflows are able to transport metals to the outskirts, but do not strongly impact the temperature and densities of gas far from the halo center.
        
        Dwarfs at intermediate masses make up a middle ground between slow and and fast quenching (see Figure \ref{case-studies-2}); {\it r398} quenches slowly over time in a way more reminiscent of the lower mass dwarfs, while {\it r244} quenches rapidly more akin to higher mass dwarfs. These dwarfs appear to exist at a critical mass where quenching changes from the explosive ``blow-outs'' found among massive galaxies to the slow heating found in lower mass dwarfs.
        
        Note that in all of these mass bins, the star-forming cases also contain an MBH. Although hosting a central MBH is typically necessary to quench a dwarf galaxy in these simulations, it is not alone sufficient in order to quench. The MBH needs to be hosted within an environment where it can accrete efficiently. In lower mass dwarfs, the MBH must undergo sustained accretion for long periods of time. To get complete quenching in higher mass dwarfs, the MBH must rapidly accrete a large amount of gas. Although mergers might help in funneling large amounts of gas toward the galaxy center, mergers are not necessary for driving quenching, even among these higher mass dwarfs.
        
\section{Caveats}

    As discussed in \citet{Sharma2020}, the primary caveat when analyzing MBHs in \Rom{} is the large seed mass of MBHs ($10^6$ M$_{\odot}$) relative to analytic estimates for direct-collapse seeding \citep{Haiman2013}. MBHs are seeded at these relatively large masses to better resolve their dynamics within the simulation \citep{Tremmel2017}. However, the Bondi-Hoyle accretion prescription scales with MBH mass which, if the MBH mass is already over-large to begin with, may drive unrealistic accretion onto the MBH. Indeed, \citet{Sharma2020} finds that overmassive MBHs in \Rom{} tend to lie at higher masses than expected from observational constraints of the $M_{\rm BH} - M_{\rm star}$ relation.
    
    Similarly, the Bondi-Hoyle accretion prescription, which appears to work for massive galaxies, may not apply to low-mass galaxies. \citet{Sharma2022} find that \Rom{} is able to reproduce observed relationships between AGN luminosity, SFR, and stellar mass among dwarf AGN, but the simulation also produces too many bright dwarf AGN relative to observations. Instead, a two-mode accretion model \citep{Churazov2005} with a dependence on mass or Eddington fraction that allows for similar amounts of growth but with a lower seed mass may better match observations of the AGN fraction.
    
    In this work, we push down to the lowest resolvable scales in \Rom{} in order to capture both MBH physics and quenching physics in dwarf galaxies. It is possible that our results are affected by numerics at the lowest dwarf masses, where the lower resolutions lead to fluffier, low resolution gas that is more susceptible to feedback. Note, however, that related analyses by \citet{Dickey2021} indicate that resolution effects may not fully explain the discrepancy in quiescent fractions between simulations and observations below $M_{\rm star}/M_\odot = 10^{9}$. They find the discrepancy in quiescent fraction compared to observations is still present when recalculating their results for the smaller, higher-resolution runs of \textsc{EAGLE} (moving from the $100$ Mpc to the $25$ Mpc boxes) and \textsc{SIMBA} (moving from the $100$ Mpc to the $25$ Mpc box). Depending on the simulation, moving to higher-resolution runs either decreased or increased the discrepancy, but never eliminated it. It is hence also possible that running higher resolutions of \Rom{} would still not solve the discrepancy in quiescent fractions.
    
    As discussed in \citet{Ricarte2019}, \Rom{} is one of several state-of-the-art cosmological simulations which have found that mergers are not the primary mode by which low- and intermediate-luminosity AGN are triggered \citep{Martin2018,McAlpine2018,McAlpine2020,Steinborn2018}. Results for \Rom{} indicate that quenching progresses through sustained growth of the stellar and MBH components. However, high-resolution simulations suggest that MBH assembly may depend dramatically on resolution \citep{Angles-Alcazar2017} -- \Rom{} may not resolve MBH accretion of small-scale, high-density gas brought in through mergers. Higher resolution zoom cosmological simulations may be required to fully disentangle the effects of numerics on MBH growth through mergers, MBH feedback, and dwarf galaxy quenching.
    
    All in all, it is likely that the AGN in \Rom{} quench isolated dwarf galaxies too efficiently, though it is unclear to what extent. While \Rom{} can reproduce local scaling relations, the resolution and/or physical prescriptions likely do not capture the true nature of accretion and feedback in low-mass galaxies. On the other hand, as shown in \citet{Dickey2021}, other cosmological simulations without MBHs in $M_{\rm star} < 10^{9} M_\odot$ dwarfs still exhibit an overabundance of isolated, quenched dwarfs. Evidently there is a need for higher resolution simulations with an emphasis on MBH growth and feedback in the lowest mass galaxies. 
        
\section{Conclusion} \label{Conclusion}
    
    We analyze quenching among isolated dwarf galaxies in the \Rom{} cosmological hydrodynamic simulation, which uniquely seeds MBHs in resolved, low mass galaxies in the early Universe. Using the mock observation package {\it orchard}, we generate mock sight-lines through the simulation box in order to measure quiescence as an observer would define it. This method allows us to make direct comparisons with observations and other mock-observed simulations from \citet{Dickey2019}. We find that:
    
    \begin{itemize}
        \item Similar to other large-scale cosmological simulations, \Rom{} predicts a greater number of isolated, quiescent dwarfs with stellar masses below $M_{\rm star}/M_\odot < 2\times10^{9}$ at $z=0.05$ than is observed (Figure \ref{qfrac-comparison}). As seen in Figure \ref{qfrac-bhs}, this discrepancy is largely explained by the presence of dwarfs with central MBHs, which exhibit $2-10$ times higher quiescent fractions relative to dwarfs without MBHs.
        \item Quiescent dwarfs form their stars then quench around $z\sim0.5-1$, as seen in Figure \ref{galaxy-growth}, though the precise quenching time depends on stellar mass.
        \item Quenching is strongly associated with energy injection into the ISM by MBHs, as seen in Figure \ref{etot-mstar}. Figures \ref{galaxy-growth} and \ref{gas-fraction} indicate that, by $z=0.05$, quiescent dwarfs have significantly less HI gas, as well as less total gas, than star-forming dwarfs. Quiescent dwarf galaxies are more likely to host an MBH that has accreted more mass over cosmic time, thereby releasing more energy in the form of AGN feedback.
        \item Dwarf galaxies across two dex in stellar mass quench at the onset of energetic black hole outflows, similar to what is seen in more massive galaxies \citep{Pontzen2017, Tremmel2019, Sanchez2021}. The most massive dwarfs in our sample (M$_{\rm star} = 10^{9.3} - 10^{10}$ M$_{\odot}$) experience rapid blowout events that significantly decrease the overall gas content of the halo. Lower mass dwarfs (below $\sim10^{9.3}$ M$_{\odot}$), on the other hand, experience a more gradual quenching process with continuous black hole activity and outflows with less extreme changes to their halo gas content.
    \end{itemize}
    
    Our simulated dwarf galaxies, along with a slew of recent observational results \citep[e.g.][]{Silk2017, Bradford2018, Penny2018, Dickey2019, Manzano-King2019}, support the idea that AGN feedback may be an important ingredient to dwarf galaxy evolution. It is likely that a large number of dwarfs host an MBH \citep[e.g.][]{Baldassare2020} and it is already known that dwarf galaxies can be an important laboratory to constrain MBH formation \citep{Volonteri2009, Volonteri2012, Greene2012}. In this work we showed the crucial impact of MBHs in creating a population of quenched, isolated dwarf galaxies. This demonstrates that low mass galaxies may also provide crucial constraints on the nature of AGN feedback. Modern cosmological simulations include a variety of different models for AGN feedback, yet these models are often created and optimized with massive galaxies in mind. The ability to model MBHs in dwarf galaxies and test against observations at low mass can provide a new way to constrain these models. Higher resolution simulations which we will perform in future work will provide even more insight into the detailed interactions between MBHs and the ISM of dwarf galaxies while new implementations of AGN feedback will be tested against observed dwarfs, as well as high mass galaxies.
    

\section{Acknowledgements}

    This material is based on work supported by the National Science Foundation under grant No. NSF-AST-1813871. MT is supported by an NSF Astronomy and Astrophysics Postdoctoral Fellowship under award AST-2001810. JMB acknowledges support from NSF AST-1812642. \Rom{} is part of the Blue Waters sustained-petascale computing project, which is supported by the National Science Foundation (awards OCI-0725070 and ACI-1238993) and the state of Illinois. Blue Waters is a joint effort of the University of Illinois at Urbana-Champaign and its National Center for Supercomputing Applications. \Rom{} initial conditions and preliminary analysis used the Extreme Science and Engineering Discovery Environment (XSEDE), which is supported by National Science Foundation grant number ACI-1548562. RSS would like to thank Nicole Sanchez, Claire Dickey, and Tjitske Starkenburg for helpful conversations and data from private communications.
    
\bibstyle{aasjournal}
\bibliography{library}
\end{document}